\documentclass[range]{ar2e}

\newcommand{\boldgamma}{\mbox{\boldmath $\gamma$}}

\newcommand{\boldzeta}{\mbox{\boldmath$\zeta$}}
\newcommand{\boldeta}{\mbox{\boldmath$\eta$}}
\newcommand{\boldmu}{\mbox{\boldmath$\mu$}}

\newcommand{\eq}[1]{Eq.~(\ref{#1})}
\def\poinc{Poincar\'{e} }
\def\bfq {{\bf q}}

\def\bfk{{\bf k}}
\def\bfp{{\bf p}}  \def\bfq {{\bf q}}\def\bfb{{\bf b}}\def\bfB{{\bf B}}
\def\bfr {{\bf r}}\def\bfR {{\bf R}}

\def\bfr{{\bf r}} \def\bfR{{\bf R}}

 \def\bfkappa {\mbox{\boldmath $ \kappa$}} 
\def\be{\begin{equation}}
 \def \ee{\end{equation}}
\def\bea{\begin{eqnarray}}
  \def\eea{\end{eqnarray}}

\usepackage{epsf}
\usepackage{epsfig}
\begin{document}

\input epsf.tex    
\input epsf.def   

\input psfig.sty

\jname{Ann. Rev. Nucl. Part. Sci...}
\jyear{2010}
\jvol{60}
\ARinfo{?}

\title{Transverse Charge Densities}

\markboth{Gerald A. Miller}{Transverse Charge Densities}

\author{Gerald A. Miller
\affiliation{Department of Physics, 
University of Washington, Seattle, WA 98195-1560, USA}}

\begin{abstract}
Electromagnetic form factors have long been used to probe the
underlying charge and magnetization densities  of hadrons and
nuclei.
Traditional three-dimensional Fourier transform methods are not
rigorously applicable for systems with constituents that move 
relativistically.
The use of the transverse charge density
 is a new, rigorously defined way to 
analyze electromagnetic form factors of
hadrons.
This review is concerned with the following issues: what is  a
transverse charge density; how is one  extracted one from elastic
scattering data; the existing results; what is the relationship with other
observable quantities; and, future prospects.
\end{abstract}

\maketitle

\section{INTRODUCTION}

A truly impressive level of experimental technique, 
effort and ingenuity has been applied
to   measuring  the electromagnetic form factors
of the proton, neutron (nucleon) and pion
\cite{Gao:2003ag,HydeWright:2004gh,Perdrisat:2006hj,Arrington:2006zm}.
These quantities are  probability amplitudes
that a given hadron 
can absorb a specific  amount of momentum and  remain in the ground
state, 
 and therefore  should supply  information about 
charge and magnetization densities.
The  text-book  interpretation of these form factors is  that their
 Fourier transforms are measurements of the charge and magnetization 
densities.
This interpretation is deeply buried in the 
 thinking of nuclear and particle physicists 
and continues to guide intuition, as it has since the days
 of the Nobel prize-winning work of Hofstadter\cite{Hofstadter:1956qs}.
 Nevertheless, the relativistic motion of the constituents of the
 system  causes the text-book interpretation 
to be incorrect \cite{Miller:2009sg}.

The preceding statement leads to several  questions, the first
 being: Is the statement correct? If correct, how fast do the
 constituents actually have to move to violate the non-relativistic
 interpretation?  Why is it that the 
 motion of the constituents and not that of the entire system that
 is relevant?
     The answers to these questions are probably 
displayed within the existing literature. However, 
 obtaining  general clear answers has been sufficiently difficult that
 posing even the first question of this paragraph would not lead to a
 unanimous  answer.  Moreover, there
 is another more important question. If the non-relativistic approach
 is not correct: What is the  correct procedure to determine
 model-independent information regarding hadronic charge and
 magnetization densities? How should we interpret the beautifully
 precise electromagnetic form factor data being produced at Jefferson 
   Laboratory, Bates, Bonn  and Mainz? 

The aim of this review is to present and clarify  answers to these
questions. The most immediate question is addressed  here. The only way to
determine model-independent information about the charge distribution
is to study the transverse charge density $\rho(b)$ \cite{Soper:1976jc}
which gives the
density  of a hadron moving at infinite momentum 
at a transverse (to the direction of rapid motion) distance $b$ 
from the
transverse center of momentum. It is only in the infinite momentum
frame that one can define such a center. 
Information about the charge density in terms of longitudinal
coordinates is not yet available in a meaningful way.

\subsection{Electromagnetic form factors are not three-dimensional
  Fourier transforms of charge densities}

Physicists are trained to believe 
 that physics is independent of the inertial
reference frame.  In non-relativistic
quantum mechanics this invariance allows us to express the wave function
as a product of 
 a plane-wave factor (that  describes the motion of the center of
 mass) with a function that depends only on internal relative
 coordinates. Thus
for  two spinless particles:
\bea\Psi(\vec{r}_1,\vec{r}_2)=e^{i\vec{P}\cdot\vec{R}}\phi(\vec{r}),
\label{sep1}\eea 
where
$\vec{P}$ is the total momentum, $\vec{R}$ is the position of the com, and
$\vec{r}=\vec{r}_1-\vec{r}_2$. Then the non-relativistic form factor
$F_{NR}(|\vec{q}|)$, which is the probability amplitude for the system
to absorb a momentum $\vec{q}$ and remain in its initial state, 
 is  given by the integral 
\bea F_{NR}(|\vec{q}|)=\int d^3r \Large{|}\phi(\vec{r})\Large{|}^2
 e^{i\vec{q}\cdot\vec{r}/2}, \label{nfrr}\eea 
if the
 masses of the two constituents are the same. One sees that the square of a
 wave function or probability density is being probed.

However the separation appearing in \eq{sep1} is not generally valid.
For
a similar but relativistic system the Minkowski-space, 
Bethe-Salpeter wave function
$\Phi(k,P)$ (where $k$ is the relative four-momentum and $P$ is the
total four-momentum) can be written in a compact form if the
interaction kernel is given by a set of
Feynman graphs~\cite{Nakanishi:1969ph,Nakanishi:1988hp}, 
using  the Nakanishi integral representation:
\begin{eqnarray}\label{bsint}
\Phi(k;P)&=&-{i\over \sqrt{4\pi}}\int_{-1}^1dz\int_0^{\infty}d\gamma
\frac{g(\gamma,z)}{\left[\gamma+m^2
-\frac{1}{4}M^2-k^2-P\cdot k\; z-i\varepsilon\right]^3},
\end{eqnarray}
where $m$ is the mass of each constituent scalar particle and $M$ is the
hadronic mass.
The weight function 
$g(\gamma,z)$ 
itself is not singular, but 
 the singularities of the BS amplitude are  reproduced by using \eq{bsint}
 \cite{Carbonell:2008tz}.
 The key feature of \eq{bsint} is that  the
covariant wave function depends explicitly on the total four-momentum
$P$. Another way of presenting the contents of \eq{bsint}
involves  the relativistic boost operator used to obtain the wave function in
a moving frame from one in which 
the hadron is at rest.
 The Nakanishi representation allows one
to perform the boost merely by changing the value of $P$.

The explicit dependence on $P$ 
dramatically influences our understanding of
form factors because  
  the initial and final hadrons  have different momentum
and  therefore different wave functions. 
The presence of different wave functions of  the initial and final
nucleons invalidates  a naive
  probability or density  interpretation. 

Assuming that only one of the scalar constituents carry charge, the
electromagnetic form factor is obtained by evaluating the impulse
approximation
triangle
diagram, with the result \cite{Carbonell:2008tz}.
\bea (P+P')^\mu F(Q^2)=i\int {d^4k\over (2\pi)^4}(P+P'+k)^\mu(k^2-m^2)
\Phi({1\over2}P-k,P)\Phi({1\over2}P'-k,P'),\label{feynf}\eea
where $P'=P+q$. 

The form factor of \eq{feynf} seems to 
 differ markedly from the usual non-relativistic
expression \eq{nfrr}, expressed in terms of momentum space wave
 functions $\tilde\phi$:
\bea  F^{NR}(|\vec{q}|)=\int {d^3k\over (2\pi)^3}
\tilde{\phi}^*(\vec{k}+\vec{q}/2)\tilde{\phi}(\vec{k}).\label{feynfnr}\eea
It is comforting  that in the non-relativistic limit,
defined by the replacement 
\bea x={k^0+k^3\over P^0+P^3}\rightarrow{1\over 2}+{k^3\over2m},\;M-2m\ll2m,\label{nrcond}\eea
 the IMF, light-front wave function becomes identical to the rest 
frame instant-form wave function \cite{Miller:2009fc}. This means 
\cite{Miller:2009sg} that
if the condition \eq{nrcond} is true for a given model, then
 \eq{feynfnr}
(and therefore \eq{nfrr})
is obtained from \eq{feynf}. In this case, one may 
extract  $\rho(\bfr)\equiv |\tilde{\phi}(\bfr)|^2$ from a 
three-dimensional Fourier transform of $F_{NR}$.
 However, the conditions \eq{nrcond} are
 not expected to be obtained for hadrons, although they are
 valid for nuclei \cite{Miller:2009sg}. Furthermore, 
 if the charged particles of a given system have different masses, one
obtains (in the non-relativistic limit for all constituents)
\bea  F^{NR}(\vec{q})=\int d^3r\sum_i \rho_i(\vec{r}) e^{-i\vec{q}\cdot\vec{r}
 m_i/M},
\label{nrdensi}\eea
where $\rho_i,m_i$ are the charge density and mass of the $i$'th
constituent, and one 
can not obtain the charge density from the form factor.

\subsection{What is a transverse charge density?}

A proper determination of a 
 charge density requires the measurement of a matrix element of a 
 density operator.
We show here that measurements of the  hadronic form factor 
directly involve the 
three-dimensional charge
density of partons 
in the infinite momentum frame,  $\hat{\rho}_\infty(x^-,\bfb)$.
Before discussing this quantity we need to provide a brief introduction to
light-cone coordinates.

In the infinite momentum frame IMF the time coordinate $ct=x^0/\sqrt{2}$
is expressed in a frame moving along the negative $z$ direction with a
velocity nearly that of light using the Lorentz transformation as the
 variable $x^+=(x^0+x^3)/\sqrt{2}$, with the usual $\gamma$ factor is
 absorbed by a Jacobean of an integral over volume
 \cite{Susskind:1968zz}.
 The
 $x^+$ variable is canonically conjugate to the minus-component of the
 momentum operator $p^-\equiv(p^0-p^3)/\sqrt{2}$. The longitudinal 
spatial variable  is  $x^-=(x^0-x^3)/\sqrt{2}$ and its conjugate
momentum is  $p^+=(p^0+p^3)/\sqrt{2}$. It is this plus-component of
momentum that is associated with the usual Bjkoren variable. The
transverse coordinates $x,y$ are written as $\bfb$ with the conjugate 
$\bfp$. In the literature, 
these transverse coordinates are sometimes written with a
subscript $\perp$, but here we shall simply use boldface to denote the transverse components of position and
momentum vectors. We shall also always use a light-front time quantization which sets $x^+$ and the plus-component of all spatial variables to zero. This means that  $x^-$ can be thought of as the longitudinal variable $-\sqrt{2}x^3.$
One extremely useful aspect of using these
variables  is that Lorentz transformations to frames moving with different 
{ transverse} velocities do not depend on interactions. These
transformations form the kinematic subgroup of the \poinc group so
 that boosts in the transverse direction are accomplished 
using the transverse component of the position operator (as in the
non-relativistic theory). 

In the IMF, the electromagnetic charge density 
$J^0$ becomes $J^+$ and
\bea \hat{\rho}_\infty(x^-,\bfb)=J^+(x^-,\bfb)=\sum_q e_q \overline{q}(x^-,\bfb)\gamma^+q(x^-,\bfb)=\sum_q e_q \sqrt{2} 
q^\dagger_+(x^-,\bfb)q_+(x^-,\bfb),\label{imfop}\eea
where  $q_+(x^\mu)=
\gamma^0\gamma^+/\sqrt{2} q(x^\mu)$, the independent part of the
 quark-field operator $q(x^\mu)$.
The 
  time variable, $x^+$ is set to  zero. Note the appearance of 
 the absolute square
  of quark field-operators, which is the signature of a true density.

We are concerned with the relationship between charge density and 
the  electromagnetic form factor $F(Q^2)$,  
determined from the current density via 
\bea F(Q^2)={\langle {p'}^+,\bfp'|J^+(0)|p^+,\bfp\rangle\over 2p^+},\label{fdef}\eea
where  the  normalization is  
$\langle {p'}^+,\bfp'| {p}^+,\bfp,\rangle
=2p^+(2\pi)^3  \delta({p'}^+-p^+)\delta^{(2)}({\bfp}'-\bfp)$.  Spin or
helicity dependence is ignored in the present sub-section.  We take 
 the momentum transfer $q_\alpha=p'_\alpha-p_\alpha$ to be
 space-like, with 
the square of the space-like four-momentum transfer $q^2=-Q^2$  
and  use  the Drell-Yan (DY) frame with 
$ (q^+=0,Q^2=\bfq^2)$. No longitudinal momentum is transferred,  so
that initial and final states are related only by kinematic
transformations. Moreover, with this condition the current operator links Fock-state components with the same number of constituents.
The  matrix element appearing in \eq{fdef} involves the combination
of creation and destruction operators: $b^\dagger b- d^\dagger d$ 
for each flavor of quark,
so that the valence charge density  is probed.
The form factor $F$ is independent of
 renormalization scale  because the 
vector current $\bar{q}\gamma^\mu q$ is conserved \cite{Diehl:2002he}. 

The spatial structure of a  hadron can be examined  using
 states that are transversely localized
\cite{Soper:1976jc,Burkardt:2002hr,Diehl:2002he}
through  a  linear superposition 
\be 
\left|p^+,{\bf R}= {\bf 0}\right\rangle
\equiv {\cal N}\int \frac{d^2{\bf p}}{(2\pi)^2} 
\left|p^+,{\bf p}  \right\rangle,
\label{eq:loc}
\ee
where 
${\cal N}$ is a normalization factor satisfying
$\left|{\cal N}\right|^2\int \frac{d^2{\bf p}_\perp}{(2\pi)^2}=1$.
Wave packet representations can be used to   avoid states 
normalized to $\delta$ functions 
\cite{Burkardt:2000za,Diehl:2000xz},
but this   leads to the
same results as using \eq{eq:loc}. 
Considering
$2p^+p^--\bfp^2=m_\pi^2>0$, (with $p^->0$)
one finds  $p^+$ must be very large because the
range of integration over $\bfp$ is large.
Using such an 
ultra-large or infinite  value of $p^+$,      maintains the interpretation
of a pion moving in the IMF with well defined  
longitudinal momentum \cite{Burkardt:2000za}. 
It is in just such an infinite momentum  that 
the parton interpretation of a hadron  is valid. 
Setting 
the  transverse 
center of momentum  
 to zero, \eq{eq:loc}, allows a meaningful
 transverse distance $\bfb$.
 A
Fock-space  parton
representation of the position of the transverse center of momentum,
$\bfR$, provides a  useful relation between longitudinal momentum
fractions and transverse positions \cite{Burkardt:2002hr}:
\bea \bfR=\sum_i\;x_i\bfb_i,\label{fock}\eea
where the sum is over the finite number of constituents in a given component.

Next  we relate the charge density  
\bea {\rho}_\infty(x^-,\bfb)={ \left\langle p^+,{\bf R}= {\bf 0}
\right| \hat{\rho}_\infty(x^-,\bfb)
\left|p^+,{\bf R}= {\bf 0}
\right\rangle
\over 
\left\langle p^+,{\bf R}= {\bf 0}
|p^+,{\bf R}= {\bf 0}
\right\rangle},\eea
to $F(Q^2)$.
In the DY frame no momentum is transferred in the plus-direction, so  that
information regarding the $x^-$ dependence of the density is not
accessible. 
Therefore we 
integrate over $x^-$, using the relationship
$q^\dagger_+(x^-,\bfb)q_+(x^-,\bfb)=
e^{i\widehat{p}^+x^-}e^{-i\widehat{\bfp}\cdot\bfb}q^\dagger_+(0)q_+(0)
e^{i\widehat{\bfp}\cdot\bfb}e^{-i\widehat{p}^+x^-},
$ 
to find
\bea
\rho(b)\equiv\int dx^-\rho_\infty(x^-,\bfb)=
\left\langle p^+,{\bf R}= {\bf 0}
\right| \hat{\rho}_\infty(0,\bfb)
\left|p^+,{\bf R}= {\bf 0}
\right\rangle/(2p^+). \label{xmint}
\eea
 The use of \eq{fdef} and the expansion 
\eq{eq:loc} leads to  
the simplification of the right-hand-side of the above equation:
\bea
\rho(b)=\int {d^2q\over (2\pi)^2} F(Q^2=\bfq^2)
e^{-i\bfq\cdot\bfb}, 
\label{rhob0}\eea
where $\rho(b)$ is termed the transverse charge density, giving the charge
density at a transverse position $b$, 
irrespective of the value of the longitudinal position or
momentum. The use of a three-dimensional coordinate space density 
$\hat{\rho}_\infty(x^-,\bfb)$ to obtain
the transverse density appeared  in Ref.~\cite{Miller:2009qu}.
 Previous derivations
\cite{Soper:1976jc,Burkardt:2002hr,Diehl:2002he,Miller:2007uy} 
used a density operator involving transverse position and longitudinal
momentum variables; see sect.~2.

If the non-relativistic (NR) limit  \eq{nrcond} is valid, one finds
that
\cite{Miller:2009sg}
\bea \rho^{NR}(b)=\int_{-\infty}^\infty dz \rho(r),\label{zint}\eea
where $\rho(r)$ is the square of the wave function 
obtained from \eq{nfrr}.  This, as well as the
integral 
over the longitudinal coordinate that appears in \eq{xmint},
 illustrates the nature of the transverse density as a 
reduction of the three-dimensional density. Moreover, 
these  integrals 
appear in the Glauber theory \cite{Glauber:1970jm} of 
high energy scattering in which scattering amplitudes 
are expressed in terms of transverse densities. A recent example is found in 
\cite{Luzum:2009sb}.

\subsection{Simple example}
It is worthwhile to use a simple example \cite{Miller:2009sg} 
in which a scalar particle $\Psi$ is
modeled as a bound state of two different scalar particles
$\xi,\chi$, with a point-like coupling such that the interaction
Lagrangian is $g\Psi\xi\chi$. Then the Bethe-Salpeter wave function is
obtained from  \eq{bsint} by replacing $g(\gamma,z)/\sqrt{4\pi}$ by
the simple coupling-constant $g$. The explicit Bethe-Salpeter wave
function can then be obtained by straightforward integration and found
to be proportional to the product of Klein-Gordan propagators.
For this model, 
the evaluation of the form factor \eq{feynf} can be performed in
three different ways. One can simply evaluate the Feynman integral,
one can take work in the IMF with $P^3\rightarrow\infty$
\cite{Gunion:1973ex}
 or one can
proceed
by first integrating over $k^-$. The resulting form factor is the same
in all three cases \cite{Miller:2009sg}. The form factor
can be expressed in terms of a
three-dimensional integral which involves a light-front wave function:
  $\psi(x,\bfkappa)$ 
\bea \psi(x,\bfkappa)\equiv g[M^2-{\bfkappa^2+m_1^2\over x}- {\bfkappa^2+m_2^2\over 1-x}]^{-1} \label{wf}
.\eea In that case
\bea F(Q^2)={1\over 2(2\pi)^3}\int d^2\bfkappa 
\int_0^1 {dx\over x(1-x)}\psi^*(x,\bfkappa+(1-x)\bfq)\psi(x,\bfkappa).
\label{2dft}\eea
The relative variables are $x,\bfkappa$  with $x=k^+/P^+$ and
$\bfkappa=\bfkappa_1-\bfkappa_2$.
The function $\psi(x,\kappa)$ can be obtained from $\Phi$ of
\eq{bsint}  by
integrating over $k^-$ \cite{Lepage:1980fj,Carbonell:1998rj,Miller:2009fc}.

\begin{figure}	
\centerline{\psfig{figure=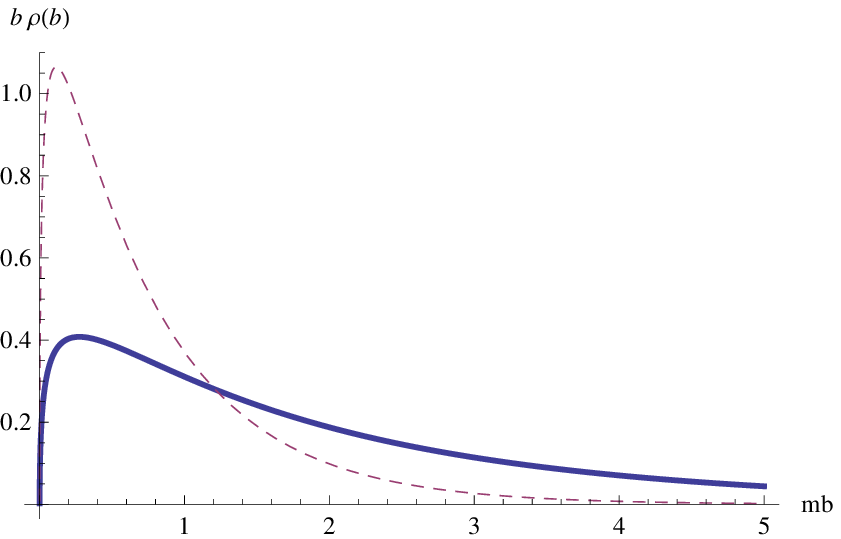,height=15pc}}
\caption{$b\rho(b)$ for the  model of \eq{tdt}.
 Solid $\epsilon=0.01$,dashed $\epsilon=0.1$.}
\label{figure1}
\end{figure}

One can obtain a relation between the form factor and a 
 coordinate space density, by expressing the light-front wave function
$\psi(x,\bfkappa)$ in terms of the canonically conjugate spatial
 variable
$\bfB=\bfb-\bfb_2$, where $\bfb$ is the transverse position operator
 of the charged constituent. Then
$\widetilde{\psi}(x,\bfB)=\int {d^2\kappa\over (2\pi)^2}
e^{i\bfkappa\cdot\bfB}\psi(x,\bfkappa),$
and
\bea F(Q^2)={1\over 4\pi}\int d^2\bfB 
\int_0^1 {dx\over x(1-x)}\vert\widetilde{\psi}(x,\bfB)\vert^2
e^{-i\bfB(1-x)\cdot\bfq}.
\label{2dfcoord}\eea
We express the form factor in terms of the position of the charged
constituent by 
using the condition that the center of transverse momentum  is set to
zero (\eq{fock}
with two components): ${\bf R}=0=x\bfb+(1-x)\bfb_2$ so that $\bfB=\bfb/(1-x)$.
Expressing the form factor in terms of ${\bf b}$  gives 
\bea F(Q^2)={1\over 4\pi}\int d^2\bfb 
\int_0^1 {dx\over x(1-x)^3}\vert\widetilde{\psi}(x,{\bfb\over 1-x})\vert^2
e^{-i\bfb\cdot\bfq}.
\label{2dfcoord2}\eea
Comparing with \eq{rhob0} allows us to identify the transverse density
as 
\bea \rho(b)=\pi\int_0^1 {dx\over
  x(1-x)^3}\vert\widetilde{\psi}(x,{\bfb\over 1-x})\vert^2,\label{tdt}\eea
which for the present  model is evaluated as 
\bea \rho(b)=
{g^2\over2(2\pi)^3}\int_0^1  dx{x \over
  (1-x)}  K_0^2(\sqrt{m^2-M^2x(1-x)}\;{b\over1-x}).\eea
The mass $M$ must be less than $2m$ for the hadron $\Psi$ to be
stable.
 We define a positive binding
fraction $\epsilon$, so that $M=2m-\epsilon m$ ($0<\epsilon<2$). Small
values of $\epsilon$ correspond to the applicability of the 
non-relativistic  limit.
The transverse densities of \eq{tdt}
 results are shown in Fig.~\ref{figure1}, for  $\epsilon=0.01 $ and 0.1.
As expected from non-relativistic intuition, the smaller value of the
binding energy corresponds to a greater spatial extent.

\section{ZOO OF NUCLEON DISTRIBUTIONS AND DENSITIES }

 The vast literature concerning 
the  distribution functions that are used to
describe nucleon structure  includes
generalized parton distributions GPDs \cite{Mueller:1998fv,Ji:1996nm,
Radyushkin:1997ki,Collins:1996fb,Ji:1998pc,Radyushkin:2000uy,Goeke:2001tz,
Diehl:2003ny,Ji:2004gf,Belitsky:2005qn,Hagler:2004yt,Boffi:2007yc}
and transverse momentum distributions 
\cite{Collins:1981uw,Ralston:1979ys,Belyaev:1988xu,Anselmino:1994gn,Mulders:1995dh,Chibisov:1995ss,Pasquini:2009eb}.
The relationship
between GPDs, transverse charge densities and TMDs is discussed in
 the present section.

We begin by discussing generalized parton distributions,  GPDs 
and follow the discussion of Burkardt
\cite{Burkardt:2002hr}.
Deep-inelastic scattering experiments allow
the determination of parton distribution functions 
(PDFs), which give the probability that quarks carry a given fraction $x$ of the nucleon longitudinal ($+$)momentum in the
infinite momentum frame (IMF).
PDFs are  the forward matrix element of 
a light-like correlation function, 
\bea 
q(x) &=&\left\langle P,S\left| 
\hat{O}_q(x,{\bf 0}) 
\right|P,S\right\rangle
\label{eq:pd}
\eea
with
\bea
& &\hat{O}_q(x,{\bf 0}) 
\equiv \int\!\! \frac{dx^-}{4\pi}
\bar{q}(-\frac{x^-}{2},{\bf 0})
\gamma^+ q(\frac{x^-}{2},{\bf 0})
e^{ix{p}^+x^-},\label{eq:0perp}
\eea
where $|P,S\rangle$ represents the wave function.
Here  we  use the light-cone gauge 
$A^+=0$.
The use of the canonical field expansion for the quark field-operators
shows explicitly  that  
PDFs give the probability that the quarks carry a longitudinal
momentum fraction $x$. We do not display  the 
scale dependence of the PDFs to simplify the notation.

Generalized parton distributions (GPDs)\cite{Mueller:1998fv,Ji:1996nm,
Radyushkin:1997ki,Collins:1996fb,Ji:1998pc,Radyushkin:2000uy,Goeke:2001tz,
Diehl:2003ny,Ji:2004gf,Belitsky:2005qn,Hagler:2004yt,Boffi:2007yc}
which describe  the scaling limit in 
real and virtual Compton scattering experiments,
are defined by allowing the momenta and spins of the initial and final nucleons
to  differ:
\bea
& &\langle P^\prime,S^\prime|
\hat{O}_q(x,{\bf 0}_\perp) 
|P,S\rangle \label{eq:gpd}\\
& &= \frac{1}{2\bar{p}^+}
\bar{u}(P^\prime,s^\prime)\left(\gamma^+  
H_q(x,\xi,t)
+ i\frac{\sigma^{+\nu}\Delta_\nu}{2M} E_q(x,\xi,t)
\right)u(P,s) 
\eea
with $\bar{p}^\mu = \frac{1}{2}\left( P^\mu
+P^{\prime \mu}\right)$ being the mean momentum
of the target,
$ \Delta^\mu = P^{\prime \mu}-P^\mu$ the four 
momentum transfer, and $t=\Delta^2$ the invariant
momentum transfer. The skewedness parameter 
$\xi = -\frac{\Delta^+}{2\bar{p}^+}$ represents 
the change in the longitudinal component of the nucleon  momentum.
 
GPDs allow for a unified description of a number of hadronic
properties. If $P'=P$, the forward limit, they reduce to conventional
parton distribution functions $H_q(x,0,0)=q(x)$. The integral over $x$
causes the $x^-$ coordinate to vanish, so that the operator
$\hat{O}_q$ is converted into  
a local current operator and results
 in the appearance of  the usual Dirac form factors of the nucleon 
\bea \sum_qe_q\int dx H_q(x,\xi,t)=F_1(t)\label{f1}\\
\sum_q e_q\int dx E_q(x,\xi,t)=F_2(t)\label{f2}.
\eea

The next step is to define impact parameter parton distributions 
\cite{Burkardt:2002hr}. One  localizes the nucleon in the
transverse direction using   the superposition  
 \eq{eq:loc} but with light-cone helicity states \cite{Kogut:1969xa}
 $|P^+,\bfp,\lambda\rangle=|P,S\rangle$.  
For a transversely localized state 
$|p^+, {\bf R_\perp}= {\bf 0_\perp},\lambda \rangle$, Burkardt's
 impact parameter dependent PDF is given by
\bea
\!\!\!\!\!\!\!\!q(x,\bfb) \equiv
\quad \left\langle p^+,
{\bf R_\perp}= {\bf 0_\perp},\lambda \left|
\hat{O}_q(x,\bfb )\right|p^+,{\bf R_\perp}
= {\bf 0_\perp},\lambda\right\rangle.
\eea
The operator $ \hat{O}_q(x,\bfb )$ is obtained by a translation of the
operator appearing in \eq{eq:0perp} in the transverse plane
\bea \hat{O}_q(x,\bfb )=
 \int\!\! \frac{dx^-}{4\pi}
\bar{q}(-\frac{x^-}{2},{\bf b})
\gamma^+ q(\frac{x^-}{2},{\bf b})
e^{ix{p}^+x^-}.\label{eq:1perp}
\eea
The function $q(x,\bfb)$ gives the probability density that a quark of
momentum fraction $x$ is located at a transverse position $\bfb$.

Burkardt \cite{Burkardt:2002hr} 
used the transversely localized states of the same helicity to show that 
$q(x,{\bf b})$ is the two-dimensional Fourier transform of the GPD $H_q$:
\bea q(x,{\bf b})=\int 
{d^2q\over (2\pi)^2}e^{-i\;\bfq\cdot\bfb}H_q(x,t=-\bfq^2).\label{ft1}
\eea 
Integration over $x$ using  \eq{f1} and \eq{rhob0}  shows that
\bea
\rho(b)=\int dx \sum_q e_q \; q(x,\bfb)=\int {d^2q\over (2\pi)^2}F_1(t=-\bfq^2)
e^{-i\;\bfq\cdot\bfb}.\label{rhoblong}\eea
This means that the transverse density can be obtained either as an integral
over $x$ of $q(x,b)$ or as an integral over $x^-$ of the three
dimensional coordinate space density, as in Sect.~1.2. 
This equality of an integral over a momentum with one over a distance
is an 
example of Parseval's
theorem. In either case, model-independent information regarding the longitudinal coordinate is not available because the probe momentum is transverse. That is $q^+=0$.

In the model of Sect. 1.3 (scalar hadron made of two scalar
constituents), 
$q(x)$ can be obtained from the integrand
of \eq{tdt} as
\bea q(x,b)={\pi\over
  x(1-x)^3}\left|\tilde\psi(x,{\bfb\over1-x})\right|^2.\label{exqxb}\eea
One sees, for large values of $x$ approaching unity, that the
transverse extent is very narrow because this corresponds to very
large values of the relative transverse separation $\bfB=\bfb/(1-x)$ for finite
values of $\bfb$. Thus the transverse spatial extent depends explicitly on the value of $x$. In particular, large values of $x$ are associated with small values of $b$.

Transverse momentum distributions TMDs are 
 another generalization  of parton distribution functions. 
These, which  contain probability distributions regarding both the 
 longitudinal momentum
 fraction
$x$ and transverse momentum $\bfk$ carried by the quarks, are given
 by 
\cite{Collins:1981uw,
Ralston:1979ys,Mulders:1995dh}
\bea \Phi_q^\Gamma(x={k^+\over P^+}, \bfk)=\langle P,S|
\int{d\zeta^-d^2{\boldzeta}\over 2(2\pi)^3}\; 
e^{i(k^+\zeta^--\bfk\cdot{\boldzeta}) }\;\bar{q}(0)\Gamma
 q(\zeta^-,\boldzeta)|P,S\rangle,\label{tmd}
\eea
where the time separation $\zeta^+=0$.
Here the quark field operators are the gauge invariant operators in
 gauges for which the gauge potentials vanish at space-time infinity
\cite{Ji:2003ak,Belitsky:2002sm}. These operators are necessary for
 non-zero transverse  spatial separations, $\boldzeta$.
 The operator $\Gamma$ is a Dirac matrix that defines
 the quark density.  

The initial and final states appearing in \eq{tmd} have the
 same momentum;  the  major difference between TMDs and
 GPDs. Another major difference is that GPDs involve densities in transverse coordinate space, but TMDs involve densities in transverse momentum space. However, one can use Wigner distributions  \cite{Ji:2003ak}
 to show that GPDs and TMDs are  obtained from different manipulations on
 the same operator. We define the
 reduced Wigner distribution as 
\bea W_q^\Gamma(\zeta^-,{\boldzeta},k^+,\bfk)={1\over4\pi}
\int d\eta^-d^2\boldeta e^{ik^+\eta^--i
\bfk\cdot\boldeta}\bar{q}
(\zeta^--\eta^-/2,\boldzeta-\boldeta/2)\Gamma 
q(\zeta^-+\eta^-/2,\boldzeta+\boldeta/2).
\eea

 GPDs are obtained by taking the matrix element of 
$W_q^\Gamma(\zeta^-=0,{\boldzeta}={\bf0},k^+,\bfk)$ between states of
 different
 momentum and then integrating over {\bf k} as in
\bea
H_q(x,\xi,t)=\langle P',S|\int 
{d^2\bfk\over
 (2\pi)^2}W_q^{\gamma^+}(\xi^-,{\boldzeta},k^+,\bfk)|P,S\rangle,
\eea
and TMDs are obtained by taking the matrix element of
 $W_q^\Gamma(\zeta^-,{\boldzeta},k^+,\bfk)$ between states of the very same
 momentum:
\bea  \Phi_q^\Gamma(x, \bfk)=\langle P,S|\int {d\zeta^-
\over
 (2\pi)^2}W_q^{\Gamma}(\zeta^-,{\boldzeta},k^+,\bfk)|P,S\rangle.
\eea

\subsubsection{$x$-Sum rules, the  connection to the equal time
  formalism, and lattice calculations}
 
The integral of a GPD over all values of $x$ as in \eq{f1} and \eq{f2}
converts the non-local bilinear appearing in \eq{eq:1perp} into a
local operator. More generally, the GPD is a correlation function of
quarks at the same light cone time $x^+=0$. The integral over $x$
gives also $x^-=0$, so that both the time $t$ and spatial
$z$ separations vanish. Thus one can compare matrix elements involving the
integral over $x$ with  related matrix elements computed using the
equal time formalism, $t=0$. Moreover, performing the $x$ integral
allows a connection with lattice QCD calculations.
 The lattice formulation is built using covariant Euclidean
space. If one sets $it$ and $z$ to 0 in a lattice calculation one
obtains a quantity that depends on transverse coordinates 
suitable for  comparison with a relevant transverse  experimentally
measured quantity. This relation  has been pointed out in connection
with transversity observables in  
Ref.~\cite{Broniowski:2009dt} and used in 
Refs.~\cite{Alexandrou:2008bn,Alexandrou:2009hs,Diehl:2005jf,Gockeler:2006zu}.

A similar connection \cite{Miller:2007ae} 
between the light-front and equal time formalism
occurs when using TMDs. A TMD \eq{tmd} gives the probability that a
quark has a three momentum characterized by $(x,\bfk)$. The relevant
matrix element involves quarks separated at the same 
 light-cone time $\zeta^+=0$.
Integration  over $x$ sets also $\zeta^-$ to 0, so  one obtains a
density  evaluated using quarks at the same time.

\section{ SINGULAR PIONIC TRANSVERSE CHARGE  DENSITY }

Understanding the  pion is  necessary to learn  how QCD
describes the interaction and existence of 
elementary particles.
  As a nearly massless excitation of the QCD vacuum with 
pseudoscalar quantum numbers, the pion  plays a central role in
nuclear and particle 
physics  as   the carrier of the longest ranged force between nucleons
and 
as a harbinger of spontaneous symmetry breaking.

Computing  the electromagnetic form factor of the pion, 
$F_\pi(Q^2)$, at asymptotically large values of $Q^2$ 
 from first principles was one of the early challenges to 
using  perturbative QCD in  exclusive processes 
\cite{Lepage:1979zb,Farrar:1979aw,Chernyak:1980dk,Efremov:1978rn}.
Such calculations have been extended 
\cite{Chernyak:1983ej,Li:1992nu,Braaten:1987yy,Miller:1994uf} by
including effects of higher order in the strong coupling constant  
and higher twist effects. The
lowest order results are about a factor of three smaller than existing
data, 
and  higher
order and 
higher twist effects are not small at currently available  values of $Q^2$ 
 \cite{Miller:1994uf}.
It is widely 
believed that at large enough values of momentum transfer $Q^2$
 the leading-order perturbative formula will be correct, 
but these large values may be very large indeed \cite{Radyushkin:1990te}.
As a result 
 there is considerable experimental interest in determining the transition 
to the region where perturbative QCD can be applied. New
measurements 
\cite{Blok:2008jy,Huber:2008id} have been performed and  more are
planned \cite{pi12}.  Here we review 
 the first   analysis  \cite{Miller:2009qu} 
that provided a model-independent  pionic transverse charge density.

Recent pion data \cite{Blok:2008jy,Huber:2008id} 
provide a very accurate measurement of the longitudinal part of the 
electroproduction cross section and the  related 
pion form factor up to a value of $Q^2 $ = 2.45 GeV$^2$.
 The result is that 
existing data for the pion form factor are
 well represented by the monopole form 
\bea F_\pi(Q^2)=1/(1+R^2Q^2/6),\label{fit}\eea
with $R^2=0.431 \;{\rm fm}^2$. 
The expression \eq{fit} allows one to determine the transverse density
from  \eq{rhob0}
with the result:
\bea \rho(b)=\frac{3 K_0\left(\frac{\sqrt{6} b}{R}\right)}{\pi  R^2},
\label{rhobt}\eea
where $K_0$ is modified Bessel function of rank zero. 
For small values of $b$ this function
diverges as $\sim \log(b)$, hence
the transverse density is singular and infinite at the origin. This
infinity is not to be ``cured'' by a renormalization procedure because
 the charge density under  consideration 
is the  matrix element of  a valence quark operator
 between eigenstates of the full Hamiltonian.      
Furthermore, divergences of  quark distribution
functions that occur at small values of Bjorken $x$ do not occur
here because transverse charge density involves the difference between
quark and anti-quark densities. 
Many field theory models, derived even before QCD was established
\cite{Goldberger:1976vd}
obtain  a form factor that corresponds to 
a divergent transverse density. 
Note also that {any}  model, such as
vector meson  dominance or holographic QCD
\cite{Brodsky:2007hb,Kwee:2007dd,Grigoryan:2007wn}  yielding
a monopole form factor has a central   density with a
logarithmic divergence. Thus holographic QCD does not supply a
representation of the soft component \cite{Radyushkin:1990te}
of the pion wave function.

Intuition regarding a possible singularity in the central
charge density  may be gained from other
examples. 
Suppose that  the non-relativistic (NR) limit in which  the quark 
masses are heavy is applicable. Then
 the pion would be a pure $q\bar{q}$ object and
the charge density would be  the Fourier transform of the form factor
(Sect.~1.1).
Given the form factor of \eq{fit} 
 the NR three-dimensional density would be  uniquely given by
\bea \rho_{\rm NR} (r) =  \frac{3}{2\,\pi \,r\,R^2}\,
e^{\frac{{-\sqrt{6}}\,r}{R}},\,\eea
where $r$ is the distance relative to the pion center of mass. This
density diverges at $r=0$.
 If one takes
$r=\sqrt{b^2+z^2}$ as demanded by the rotational invariance of the
 non-relativistic wave
function, 
then one finds from \eq{zint} that the non-relativistic transverse
charge density takes the form of 
$\rho(b) $ of \eq{rhobt}.

The  divergence of the central transverse charge density encountered
 here may be the consequence  of using a simple
parametrization, presently  untested  by measurements at  larger
 values of $Q^2$. Thus  we  examine 
other  approaches. Consider first 
perturbative QCD (pQCD),
which provides a prediction \cite{Lepage:1979zb}-\cite{Efremov:1978rn}
for asymptotically large values of $Q^2$  that
\bea { \lim}_{Q^2\to\infty}F_\pi(Q^2) =16 \pi \alpha_s(Q^2)f_\pi^2/Q^2,\eea  
with the pion decay constant $f_\pi=93 $ MeV, and  in leading order:
\bea  \alpha_s(Q^2)={4\pi\over( 11-{2\over 3}n_f)\ln {Q^2\over
    \Lambda^2}},\label{pqcd}\eea
with $n_f$ the number of quarks of  mass smaller  than $Q$ and $\Lambda$
is a parameter fixed by data. The  $\log Q^2$ term
in the denominator does not lead to a non-singular behavior of $\rho(b)$
for small values of $b$ because  
 $\rho(b)\sim \log\log{1\over b}$ and 
 the pQCD form      factor corresponds to a singularity at
short distance. 
This singularity would arise in any model form factor even one  based on
sum rules {\it e.g} \cite{Radyushkin:1990te}
that  approaches  the pQCD asymptotically.

Chiral quark models (see the review   \cite{Broniowski:2008tg})
present other  models \cite{RuizArriola:2003bs,Broniowski:2003rp} 
of transverse charge densities that are
singular at the center as $\log b$ because 
 the pion form factor takes the monopole form.
It is nevertheless  interesting to note that each and every 
 observable quantity, including $f_\pi$ and structure functions, is 
computed to be finite.

\begin{figure}
\centerline{\psfig{figure=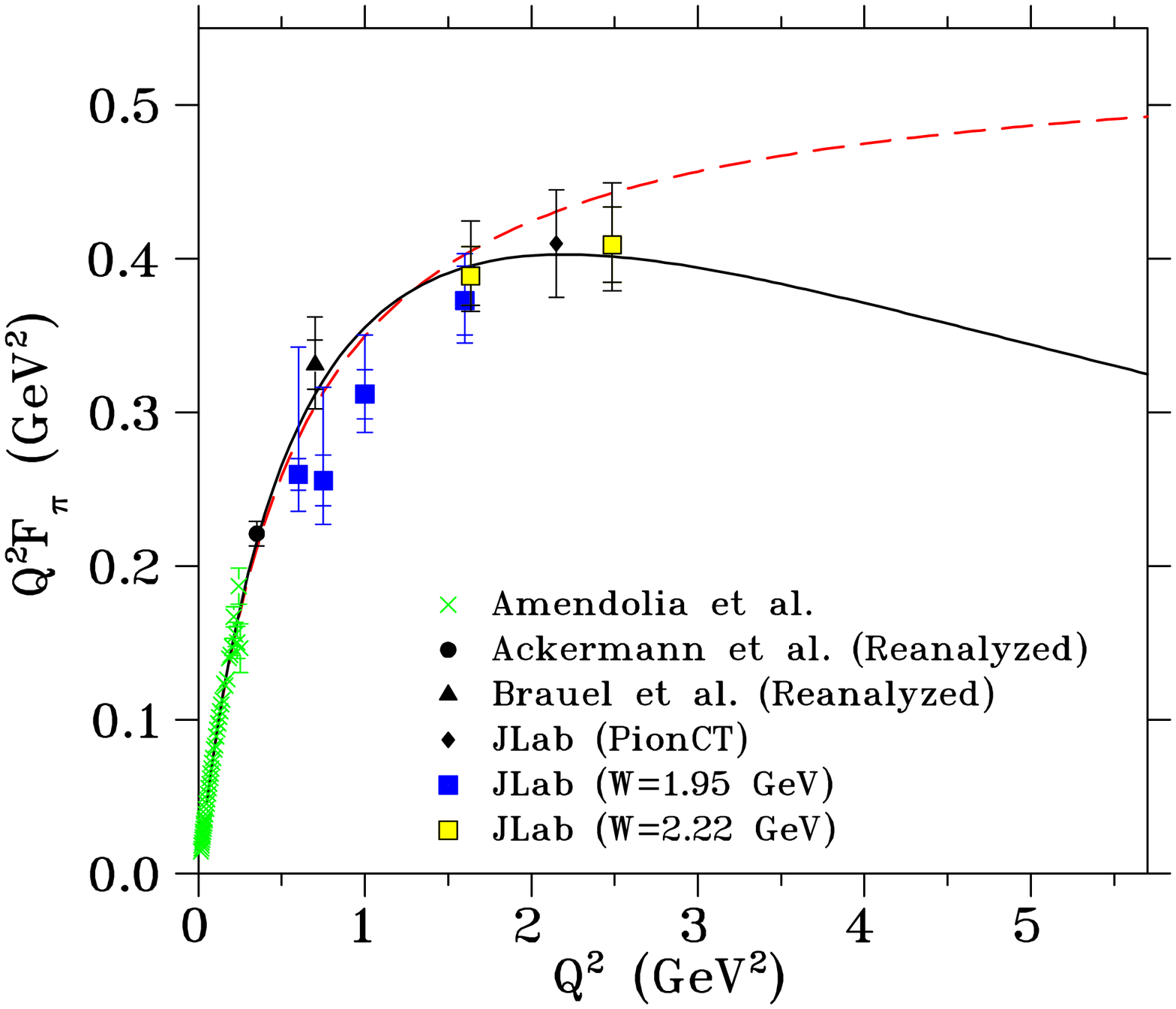,height=15pc,angle=90}}
\caption{Pion form factor data as
plotted in \cite{Huber:2008id}.  The data labeled Jlab are from
  \cite{Huber:2008id}.
The data  Brauel {\it et al.} \cite{Brauel:1979zk}  and that of
Ackermann {\it et al.} \cite{Ackermann:1977rp} have using the method
  of   \cite{Huber:2008id}. The Amendola data {\it et al.} are from \cite{Amendolia:1986wj}
The data point labeled PionCT is from \cite{Horn:2007ug}.  The (red) dashed curve uses the monopole fit \eq{fit} and the (black) solid line  the constituent
  quark model of \cite{Hwang:2001hj}.  $b\rho(b)$ for the  model of \eq{tdt}.
Reprinted from
  Ref.~\cite{Miller:2009qu}
with permission of the APS.}
\label{figure2}
\end{figure}

\begin{figure}\label{figure3}
\unitlength1.cm
\centerline{\psfig{figure=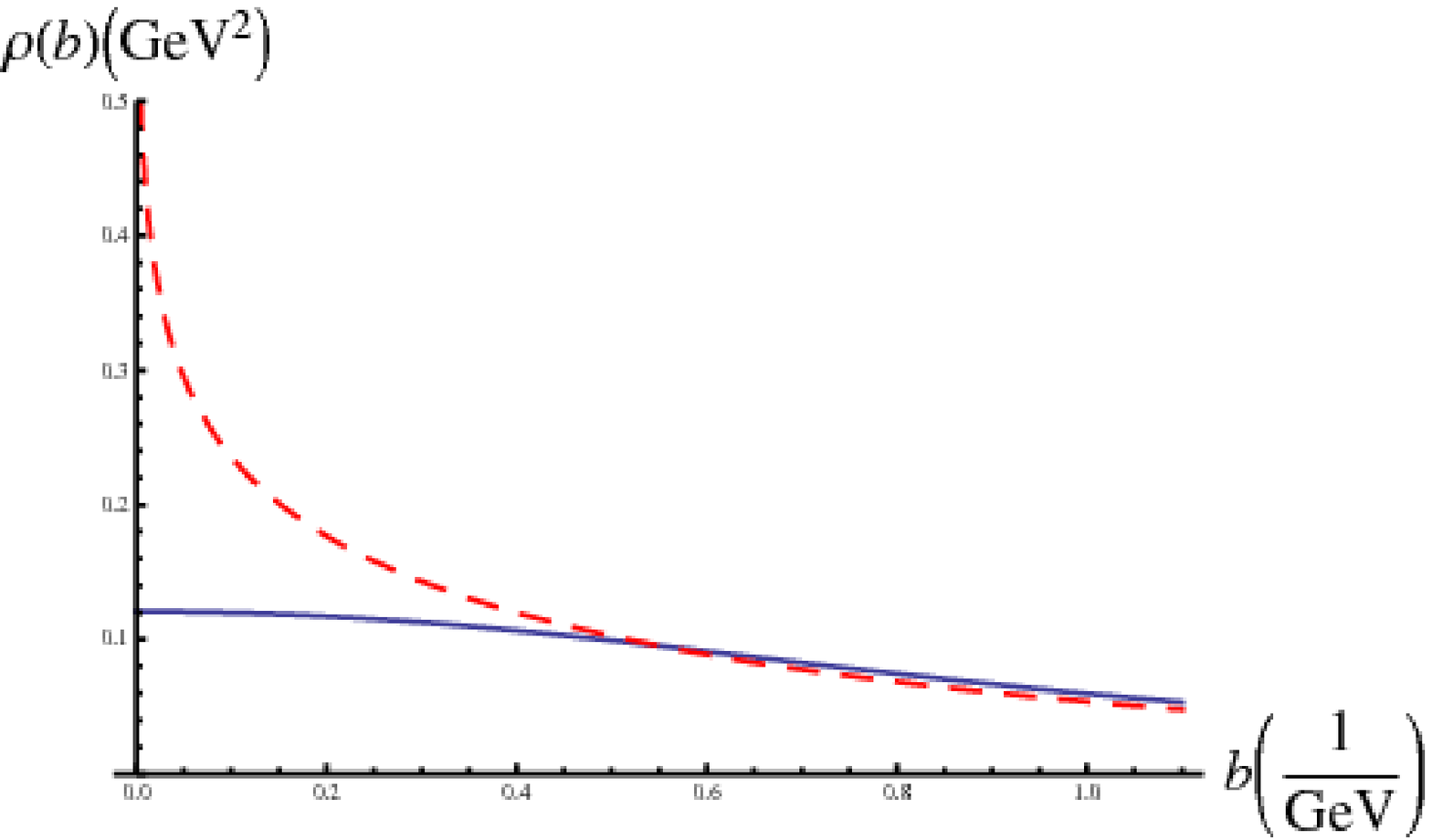,height=15pc}}
\caption{ $\rho(b)$  corresponding to  the two models
  shown in Fig.~
\ref{figure2}. The (red) dashed curve represents the transverse
  density obtained from  the monopole fit and the (blue) solid-line is
  obtained using the relativistic
constituent
  quark model of\cite{Hwang:2001hj}. Reprinted from
  Ref.~\cite{Miller:2009qu}
with permission of the APS. }
\end{figure}

Gaussian models with generalized parton distributions 
$H(x,0,Q^2)$  (see \eq{f1}) dominated by behavior
near $x=1$  present a set of examples that also 
 yield a form factor with a $1/Q^2$ asymptotic behavior,
and have a impact parameter distribution that is well behaved at each
value of $x$
 for all $b$. The 
key asymptotic features are captured in the  simple formula 
\cite{Radyushkin:1998rt,Burkardt:2003mb}:
$ 
H(x,0,Q^2)_{x\to 1}=(1-x)^{n-1}e^{-a(1-x)^n Q^2},$ with $ n>2$
so that 
$q(x,b)_{x\to1}={1\over2\pi a(1-x)}e^{-b^2/(4a(1-x)^n)}$.
This form shows that $q(x,b)$ is well behaved for all values of $b$
and for each value of $x$, but
the integral over $x$ contains a logarithmic divergence. 

Relativistic light-front constituent quark models
\cite{Chung:1988mu,Frederico:1992ye,Hwang:2001hj} 
 that describe  existing form factor data have a
non-singular central  charge density. 
These models can be  most simply derived \cite{Frederico:1992ye} by
using the impulse approximation (evaluating  the
triangle diagram).
 One starts by evaluating the integral over the 
minus component of the  loop momentum $k^\mu$, 
and then cuts off the remaining integral over $x=k^+/p^+,k_\perp$
using a phenomenological wave function.  
The
wave function 
of Ref.~\cite{Hwang:2001hj} is chosen to be  a power-law form, 
and the resulting 
model  describes
 the existing form factor data in  the space-like 
region, $f_\pi$, and the transition form factor $f_{\pi\gamma}$ in
which a virtual photon transforms a real pion into a real photon.
The model  form factor of   \cite{Hwang:2001hj} and the monopole fit
of \eq{fit} are shown along with the measured data in
Fig.~\ref{figure2}.
Both the fit and the model provide a good fit to the data, 
but present very different
predictions for larger values of $Q^2$  where measurements remain to
be done. The corresponding versions of  $\rho(b)$ are shown in
Fig.~\ref{figure3}.
The singularity of  \eq{rhobt} is manifest  as a rapidly rising
function as $b$ approaches zero, but  the relativistic constituent quark
model provides a smooth function $\rho(b)$.
The planned experiment \cite{pi12} aims to achieve results up to 
$Q^2=6\;$ GeV$^2$. This should be large enough to resolve the
differences between the model \cite{Hwang:2001hj} and the monopole
fit, 
or rule both out. However, 
this  will probably  not be sufficient to reach the
regime where pQCD might be valid \cite{pi12}.  
The model \cite{Hwang:2001hj} 
represents one useful
 phenomenology, but  
other interesting models exist. 
Computing the  transverse density in those  models 
would be useful.

\section{ NUCLEON TRANSVERSE CHARGE DENSITIES}

In previous sections we have emphasized  that
a proper determination of a nucleon 
 charge density requires a 
relation with 
the square of a field operator, and that this
can be achieved through the transverse charge density
\cite{Soper:1976jc,Burkardt:2002hr,Diehl:2002he,Miller:2007uy,Carlson:2007xd},
$\rho(b)$, 
           given by \eq{rhoblong} . We review
 the first 
 phenomenological analysis of existing data that  determined 
$\rho(b)$ for the neutron and proton \cite{Miller:2007uy}.
 The  neutron results 
contradict the ancient idea,  obtained  from
both meson-cloud and gluon-exchange  models 
\cite{Thomas:1981vc,friar72,Carlitz:1977bd,Isgur:1980hh} regarding the 
positive value of  the non-vanishing charge density 
at the center of the neutron.  The meson-cloud idea  is that 
the neutron sometimes undergoes a spontaneous quantum 
fluctuation into  heavy proton surrounded by a negatively
charged pionic cloud,  leaving  
 a positive central density\cite{Thomas:1981vc}.
The quark-model mechanism 
\cite{friar72,Carlitz:1977bd,Isgur:1980hh}
involves  the repulsive nature of the  one gluon
exchange interaction between two  $d$ quarks.

These well-motivated physical considerations  concern statements about the
three-dimensional Fourier transform of the neutron's electric form
factor
$G_E^n(Q^2)$. However, such a transform is not the charge density. The
only model-independent charge density is  $\rho(b)$.

To proceed, recall   the     definitions of the  form
factors. With 
 $J^\mu(x)$ as the electromagnetic current operator, in units of the proton 
charge, the nucleon form factors are given by 
\bea
\langle p',\lambda'| J^\mu(0)| p,\lambda\rangle =\bar{u}(p',\lambda')\left(\gamma^\mu F_1(Q^2)+i{\sigma^{\mu\alpha}\over 2M}q_\alpha F_2(Q^2)
\right) u(p,\lambda),\eea
where the momentum transfer 
$q_\alpha=p'_\alpha-p_\alpha$ is taken as space-like,  with $q^+=0$,
 so that 
$Q^2\equiv -q^2=\bfq^2>0.$ The nucleon polarization states are chosen to be those of 
definite light-cone helicities $\lambda,\lambda'$ 
\cite{Kogut:1969xa}
The charge (Dirac) form factor is $F_1$, normalized such that $F_1(0)$ is the
nucleon charge, and the magnetic (Pauli) form factor is $F_2$, 
normalized such that $F_2(0)$ is the
anomalous magnetic moment. The Sachs form factors\cite{Sachs:1962zzc}
\bea
G_E(Q^2)\equiv F_1(Q^2)-{Q^2\over 4M^2}F_2(Q^2),\; G_M(Q^2)\equiv F_1(Q^2)+F_2(Q^2),\label{sachsff}\eea
were introduced so as  to
 provide an expression for the electron-nucleon cross section
(in the one photon exchange approximation)
 that depends on the quantities $G_E^2$ and $G_M^2$, but not
the product $G_E\;G_M$.
In  the Breit frame, in which 
$\bfp=-\bfp'$,  $G_E$ is the nucleon helicity flip matrix element of
$J^0$. 
Furthermore, 
 the scattering of neutrons from the electron cloud of atoms  measures
the derivative $-d G_E(Q^2)/dQ^2$ at $Q^2=0$, 
 widely interpreted as six times
the mean-square charge radius of the neutron.  However,
any probability or density interpretation of $G_E$ 
is spoiled by a non-zero  value of $Q^2$, 
no matter how small. This is because
the  momentum difference between  the
initial and final states 
appears via the use of  derivatives 
of momentum-conserving delta functions in the moments computed by
Ref.~\cite{Sachs:1962zzc}.
 Any attempt to analytically incorporate relativistic corrections in a 
$p^2/m_q^2$ type of expansion  would be doomed, by the presence of the 
quark mass, $m_q$, to  be model-dependent. This is explained more
thoroughly in Ref.~\cite{Rinehimer:2009yv}.

We
exploit \eq{rhoblong} by  using measured form factors
to determine $\rho(b)$. 
Recent parametrizations 
\cite{Bradford:2006yz,Kelly:2004hm,Arrington:2003qk,Arrington:2007ux,Alberico:2008sz}
of $G_E$ and $G_M$ are very useful, so we use \eq{sachsff} to
obtain $F_1$ in terms of $G_E,G_M$. Then
$\rho(b)$ can be expressed as a simple  integral
of known functions:
\bea \rho(b)= \int_0^\infty\; {dQ\;Q\;\over  2 \pi}J_0(Q b) {G_E(Q^2)+\tau G_M(Q^2)\over 1+\tau},\label{use}\eea
with $\tau={Q^2\over 4M^2}$ and $J_0$ a cylindrical Bessel function.

\begin{figure}\label{figure4}
\centerline{\psfig{figure=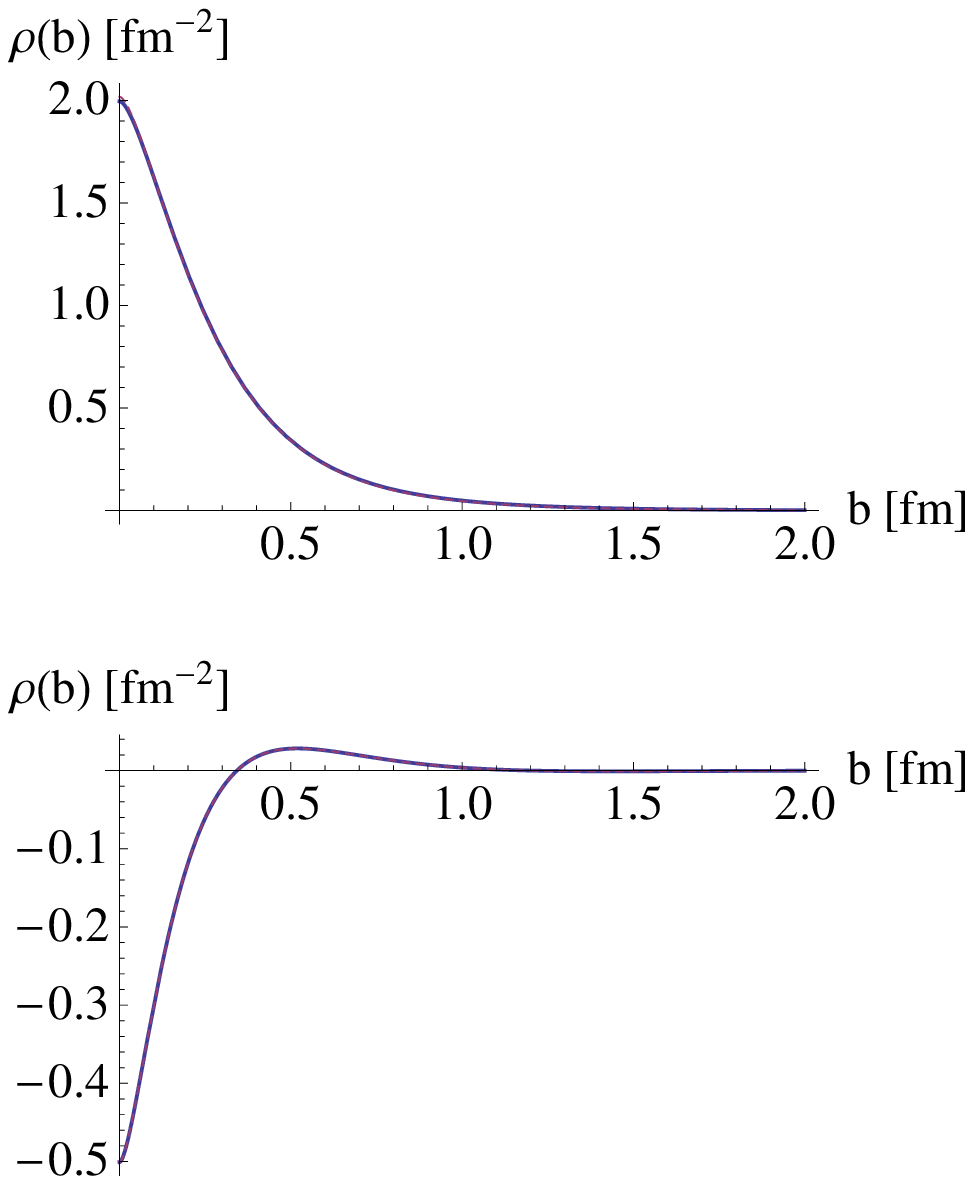,height=20pc}}
\caption{Nucleon $\rho(b)$  Upper panel: proton transverse charge density. 
Lower panel: neutron transverse charge density. 
These  densities are obtained using
 the  parametrization of \cite{Alberico:2008sz}. }
\end{figure}
A straightforward application of \eq{use} to the proton 
using the parametrizations of Ref.~\cite{Alberico:2008sz}
yields the results shown in the upper panel of  Fig.~\ref{figure4}.
 The curves obtained using the two different parametrizations
 overlap.
 Furthermore, there seems to be 
negligible sensitivity 
to form factors at very high values of $Q^2$ that are currently
unmeasured. 
The density is peaked 
at low  values of $b$, but has a long positive tail, suggesting  a long-ranged, positively charged 
  pion cloud.

The neutron  results 
are shown in the lower panel of 
 Fig.~\ref{figure4}.
 The curves obtained using the two different parametrizations seem to overlap, 
The surprising result
is that the central 
neutron charge density is negative. 
  The values of the integral of \eq{use} are somewhat sensitive to the
regime $8<Q^2<16 $ GeV$^2$ for which $G_E$ is  as yet   unmeasured. 
About 30\% of the value of 
$\rho(0)$ arises from this region. 
That 
$\rho(b=0)<0$ was confirmed in 
Refs.~\cite{Pasquini:2007iz,Carlson:2007xd,Boffi:2009sx,Hwang:2008gh}.

\begin{figure}\label{figure5}
\centerline{\psfig{figure=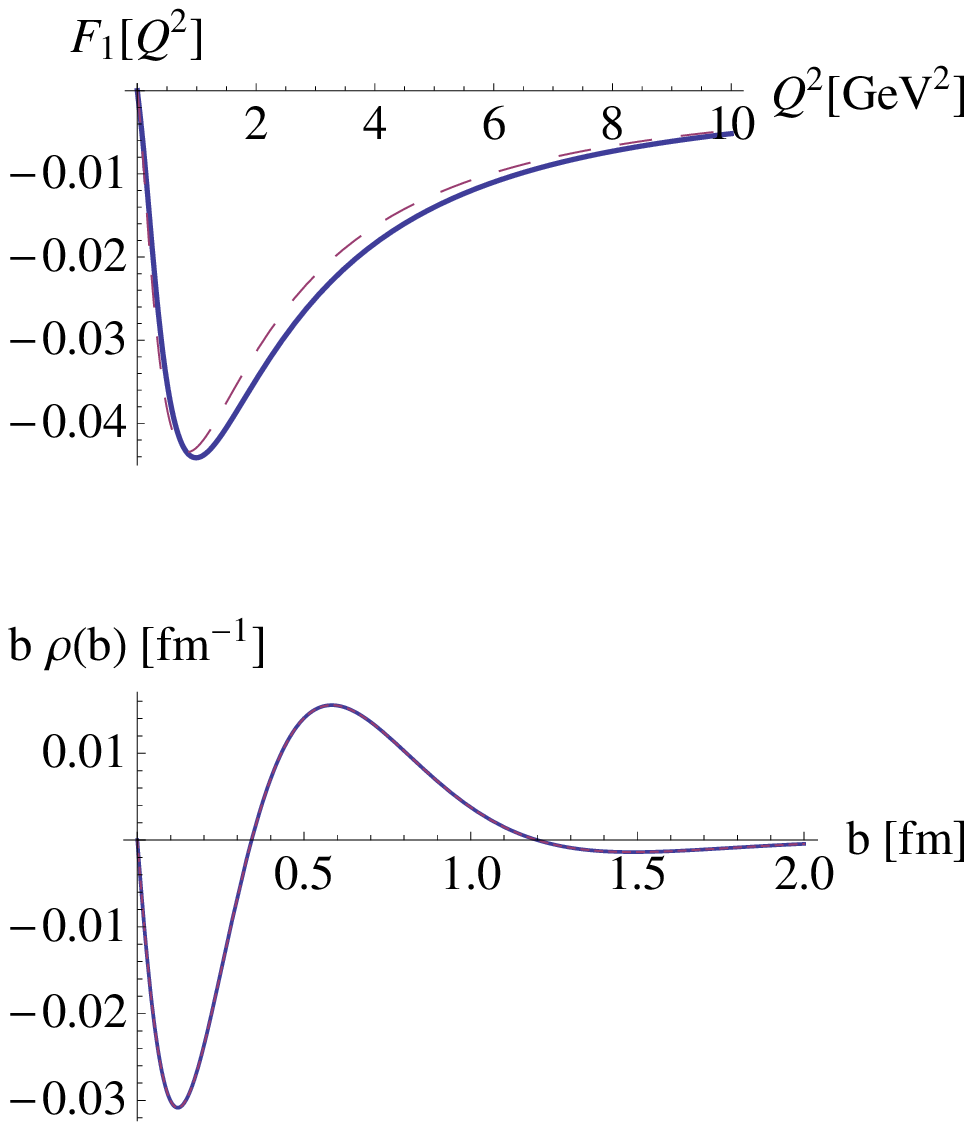,height=15pc}}
\caption{Neutron $F_1$ and $b\rho(b).$ Upper panel: $F_1(Q^2)$. 
Lower panel:   $b \rho(b)$ . The
solid curves are  obtained using Fit 1 and  and the dashed curves with
Fit 2 of  \cite{Alberico:2008sz}.  }
\end{figure}
The negative central density deserves further explanation. See 
Fig.~\ref{figure5}
in which 
the upper panel  shows $F_1$ for the neutron from 
  two  parametrizations of Ref.~\cite{Alberico:2008sz}.
In both
cases
 $F_1$ is  negative (because of the dominance of the $G_M$ term of \eq{use})
for all  values of
$Q^2$. This along with taking $b=0,\;J_0(Qb)=1$ in \eq{use} 
leads immediately to the central negative result. The long range
structure 
of the charge density is  captured by displaying the quantity
$b\rho(b)$ in the lower panel of Fig.~\ref{figure5}. 
At very large distances from the center,  $b\rho(b)<0$    
suggesting the existence of the  long-ranged pion cloud. 
Thus the neutron transverse charge density has an unusual behavior in
which the positive charge density in the middle is sandwiched by
negative charge densities at the inner and outer reaches of the
neutron.
A simple
model in which the neutron fluctuates into a proton and a $\pi^-$ 
parametrized to  reproduce the negative-definite nature of the
neutron's
$F_1$ \cite{Rinehimer:2009sz} reproduces the negative transverse central
density. In this case, the negative nature arises from pions that
penetrate to the center. The change from the nominal positive value
obtained from $G_E$ can be understood as originating in the boost to
the infinite momentum frame \cite{Rinehimer:2009yv}.

One can gain information about the individual $u$ and $d$ quark
densities by invoking charge symmetry (invariance under a  rotation by
$\pi$ about the $z$ (charge) axis in isospin space) 
\cite{Miller:1986mk,Miller:1990iz,Miller:1997ya,Miller:2006tv} and
neglecting effects of $s\bar{s}$ pairs \cite{Acha:2006my}.

Model independent information about nucleon structure 
is
obtained,
with the  particular surprise that  the central
density of the neutron is negative.
Future measurements
of neutron electromagnetic form factors could 
render the present results more precise, or potentially modify
them considerably. Obtaining a quantitative 
and intuitive understanding of these results presents a challenge
to lattice QCD and to model builders. 

\subsection{Meaning of central density $b=0$}
The transverse coordinate $b=0$ is the transverse center of the
nucleon. It is of interest to note that, in the infinite momentum
frame,  this position is also the
center of the nucleon. In  that frame the nucleon has no
longitudinal extent. This can be seen by considering a light cone 
wave function $\psi{(x,\bfkappa)}$. The canonically conjugate coordinate to
$x=k^+/P^+$ is $P^+x^-$, \cite{Pirner:2004qd}
so that the coordinate-space wave function
would depend on the product $x^-P^+$. Thus when $P^+$ is infinite,
the
coordinate-space wave function will 
 vanish unless $x^-=0$.  This means that in  the infinite momentum frame, the
 longitudinal density appears as a delta function and  the position
 $b=0$ is the true nucleon center \cite{Miller:2009sg}.

\subsection{Inclusive-exclusive Connection 
Interpretation of the negative neutron charge density}
Generalized parton distributions (GPDs) contain
information about the longitudinal momentum fraction $x$ as well as the
transverse position $b$.  Information regarding these  dependence's is
obtained from experiment via   GPDs that reproduce both deep inelastic
scattering and elastic scattering data.  Miller \& Arrington 
 \cite{Miller:2008jc}
used  this inclusive-exclusive
connection to interpret the central neutron charge density, with the
finding that the center of the neutron is dominated by  negatively
charged $d$ quarks. 
Their argument is reviewed here.

The quantities  $q(x,b)$ 
are not measured
directly, but have been obtained from models that incorporate fits to parton
distributions and electromagnetic nucleon form
factors
\cite{guidal:2004nd,diehl:2004cx,ahmad:2006gn,tiburzi:2004mh}.
 Form
factor sum rules \eq{f1}, \eq{f2} at zero skewness are exploited  to model
valence quark GPDs, $H^q_v\equiv H^q-H^{\bar{q}}$.
This yields the net contribution to the form factors from quarks and
anti-quarks. Possible effects of strangeness are
neglected in these fits.

Each parametrization  used 
\cite{guidal:2004nd,diehl:2004cx,ahmad:2006gn} incorporates 
the Drell-Yan-West \cite{Drell:1969km,West:1970av}
relationship between the behavior of the
structure function $\nu W_2(x)$ function near $x=1$, measured in
inclusive reactions  and the behavior of the electromagnetic
form factor at large values of $Q^2$, measured in the exclusive
elastic scattering process. In particular, for a system of
$n+1$ valence quarks, described by a power-law wave function
\bea
\lim_{x\rightarrow1}\nu W_2(x)=(1-x)^{2n-1} \to
\lim_{Q^2\rightarrow\infty}F_1(Q^2)=\frac{1}{Q^{2n}}.
\eea   
The  value of $n$ that defines the high-$x$
behavior of the structure function also defines the high-$Q^2$ behavior
of the form factor. This relation associates 
 the behavior of large values of $x$
with large momentum transfers, $Q^2=\bfq^2$, which in turn corresponds to small
values of $b$.  There is a further connection between large values of
$x$ and small values of $b$ which emerges from \eq{fock} and was seen
in the example of Sect.1.3 (and \eq{exqxb}). If a single quark carries
momentum fraction $x$ very near unity, only one term in the sum of
\eq{fock} survives and the restriction of this term to zero, causes
the corresponding transverse coordinate to vanish. This  interesting
connection 
between different components of the  position and momentum vectors 
 is not associated with the uncertainty principle.

Given that the $d$ quark dominates the large $x$ quark-distribution
  function of the
  neutron,
  see {\it e.g.} \cite{Pumplin:2002vw}, and that 
 the central charge density of the neutron is negative, it is natural to conclude that the central charge density of the neutron arises from the predominance of  $d$ quarks at the center. This is shown in 
 Fig.~6. 
One sees that for large values of $x$, $b$ must be small to have a
 non-zero value of $\rho_n(x,b)$, and these non-zero values are negative.

\begin{figure}\label{figure6}
\begin{picture}(14,8.2)(2.5,-1.9)
\includegraphics{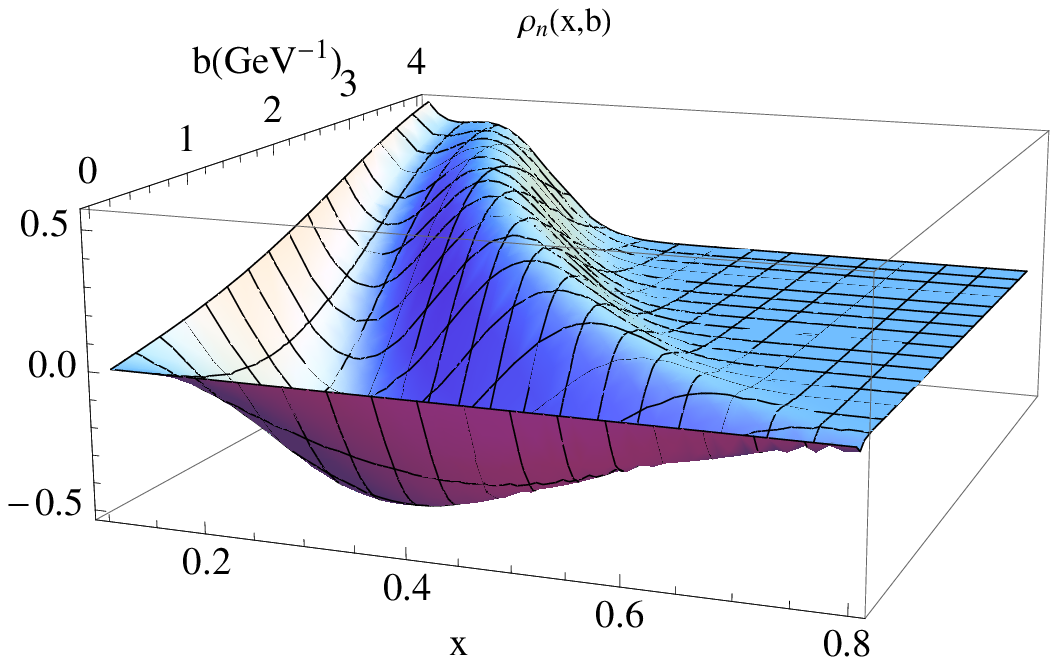}
\end{picture}
\caption{Neutron impact parameter charge distributions,
${\rho}_n(x,b)$ as a function of $b$. 
}
\end{figure}
\subsection{Magnetization Density}

We now try use transverse densities to obtain the nucleon magnetization density in the infinite momentum frame 
\cite{Miller:2007kt} 
in terms of a magnetization density. This quantity  is closely related to the transverse charge density for a polarized nucleon obtained by
Carlson and Vanderhaeghen \cite{Carlson:2007xd}.
 Our starting point is    
the relation that $\boldmu\cdot{\bf B}$ is the matrix element of
$\vec{J}\cdot\vec {A}$ in a definite state, $\vert X\rangle$.

Take the rest-frame magnetic field
to be a constant vector in the $1$ (or   $b_x$) 
  direction, and the corresponding vector
potential as 
${\bf A}= B b_y \hat{{\bf z}}$. Then  consider the
system in a 
frame in which the plus component of the momentum
approaches infinity so that ${\bf J}\cdot {\bf A}\rightarrow
J^+A^-.$
The anomalous magnetic
moment may be extracted 
by taking the matrix element of $\boldmu\cdot\bfB$ in the state
\bea \vert X\rangle\equiv \frac{1}{\sqrt{2}}\left[\left|p^+,{\bf R}= {\bf 0},
+\right\rangle+\left|p^+,{\bf R}= {\bf 0},
-\right\rangle\right],\eea
where  $\left|p^+,{\bf R}= {\bf 0},+\right\rangle $ represents a
transversely localized state of definite $P^+$ and light-cone helicity. 
The state $\vert X\rangle$ \cite{Burkardt:2002hr}
may be interpreted as that of
  a transversely polarized target \cite{Burkardt:2005hp}. 
The resulting anomalous 
magnetic moment $\mu_a$  is given by \cite{Miller:2007kt} 
\bea \mu_a=
{\langle X\vert\int d^2b \;b_y\;{q}_+^\dagger(0,\bfb)q_+(0,\bfb)\vert X\rangle},\label{mua}\eea
where  $q(x^-,\bfb)$ is a quark-field operator, and $q_+=\gamma^0\gamma^+q$.
This matrix element  is the anomalous magnetic moment
because the use of the transversely localized state $|X\rangle$ and
the
 factor $b_y$ suppresses the Dirac contribution \cite{Burkardt:2005hp}.

In Ref.~\cite{Miller:2007kt}  $\mu_a$ was evaluated by using
 Burkardt's \cite{Burkardt:2002hr}  impact parameter distribution 
and then integrating by parts. The result was 
 \bea \mu_a={1\over2M}\int d^2b \;\rho_M(b), \label{resultm}  \eea
where
\bea
\rho_M(b)=\int\frac{d^2q}{(2\pi)^2} F_2(t=-\bfq^2)e^{-i \bfq\cdot
 \bfb},
\label{mdensx}\eea
and the subscript $M$ denotes the
anomalous magnetic moment.

The expression (\ref{mdensx})
has an appealing simplicity. However, the directly-obtained 
\eq{mua} can also be evaluated immediately.
We pursue this here to find 
\bea \mu_a= {-1\over2M}\int d^2b \;b_y{\partial\rho_M(b)\over\partial
  b_y},\label{resultm1}  
\eea
so that the quantity $-b_y{\partial\rho_M(b)\over \partial b_y}\equiv 
\tilde{\rho}_M(\bfb)$   also
has an interpretation as an anomalous  magnetization density. 
The two integrals appearing in \eq{resultm} and \eq{resultm1} the same
value.
Ref.~\cite{Miller:2007kt}  rejected the use of $\tilde{\rho}_M(\bfb)$
as a magnetization 
density because of the  appearance of an explicit direction
$y$. However, 
this direction has a general interpretation as the transverse direction
orthogonal 
to that of the transverse magnetic field. We evaluate
$\tilde{\rho}_M(\bfb)$ 
in terms of $F_2(Q^2)$ to find 
\bea\widetilde{\rho}_M(\bfb)=\sin^2\phi 
\;b \int _0^\infty{q^2\;dq\over 2\pi}J_1(qb)F_2(q^2),\label{rhotilde}\eea
where $\phi$ is the angle between the direction of $\bfb$ and that of
the transverse magnetic field, which is also the direction of the
nucleon polarization. Thus the physical direction of the polarization
or magnetic field provides a definite spatial direction. 
Indeed the magnetization density $\widetilde{\rho}_M(\bfb)$ is largest in
 directions perpendicular to the direction of the nucleonic
 polarization (or magnetic field),
as  shown in Fig.~7. 
The largest values occur for $\phi=\pi/2$, and the magnetization
 density peaks at about 0.5 fm.
 Furthermore $\widetilde{\rho}_M(\bfb)$  vanishes if $b=0$ or if
 $\phi=0$. These features are  in accord with the expectations of
 classical physics. 
A current in the $z$ direction causes a magnetic dipole density $\sim
 \bfr\times\vec{J}$
 in the $x-$direction for positions $\bfr$ along the $y$-direction.  
 Therefore we conclude here that the quantity $\widetilde{\rho}_M(\bfb)$ 
is the preferred expression for the magnetization density.

Hoyer \& Kurki have computed the transverse density of the electron \cite{Hoyer:2009sg}.
Using their expression for the electron $F_2$ in \eq{resultm1} leads
to the correct result for the electron's anomalous magnetic moment,
$\mu_a^e$. Furthermore the presence of the $\sin^2\phi$ term is
consistent 
with their physical interpretation of the sign of $\mu_a^e$ 
as caused by currents in the positive (negative)  $z-$ 
direction for positive (negative)
values of $y$. Computing the cross product 
between $\bfb$ and those currents gives  magnetization in the
$x$-direction for both positive and negative values of $y$.

To better understand 
 this new  magnetization density, consider the moments of both densities.
Define the $n$th moment of $\rho_M$ as
\bea \langle b^{2n}\rangle_M\equiv \int d^2b b^{2n}\rho_M(b).\eea
These quantities are related to the $n$th derivative of $F_2(q^2)$. Similarly the $n$th moment of $\widetilde{\rho}_M$ is given by
\bea \langle b^{2n}\rangle_{\widetilde{M}}\equiv -\int d^2b
b^{2n}b_y{\partial\rho_M(b)\over \partial b_y}.\eea  Then use of
\eq{mdensx} and 
 integration by parts shows that
\bea\langle b^{2n}\rangle_{\widetilde{M}}=(n+1)\langle b^{2n}\rangle_M
.\eea
The $n=0$ moment corresponds to the anomalous 
magnetic moment which is the same for the two densities. 
The case $n=1$ which defines the mean-square magnetic radius is more interesting.
Ref.~\cite{Miller:2007kt}  showed that $\langle b^2 \rangle_M$ was
slightly larger than  $\langle b^2 \rangle_{Ch}$ (which is obtained
from $F_1$).  
However  $\langle b^2 \rangle_{\widetilde{M}}$ is
 twice as large as  $\langle b^2 \rangle_M$ and therefore is much larger than
 $\langle b^2 \rangle_{Ch}$!  This clearly shows that the proton's
magnetization density extends much further than its charge density, a
conclusion obtained in Ref.~\cite{Miller:2007kt}. This is surprising
because it contradicts simple naive intuition gained from the rapid fall of
the ratio $G_E/G_M$ with increasing momentum transfer
 \cite{Perdrisat:2006hj}, if one assumes that $G_E$ ($G_M$)
is related to the spatial extent of the charge (magnetization) density.

\begin{figure}\label{figure7}
\unitlength1.cm
\begin{picture}(14,8.2)(.5,.4)
\includegraphics{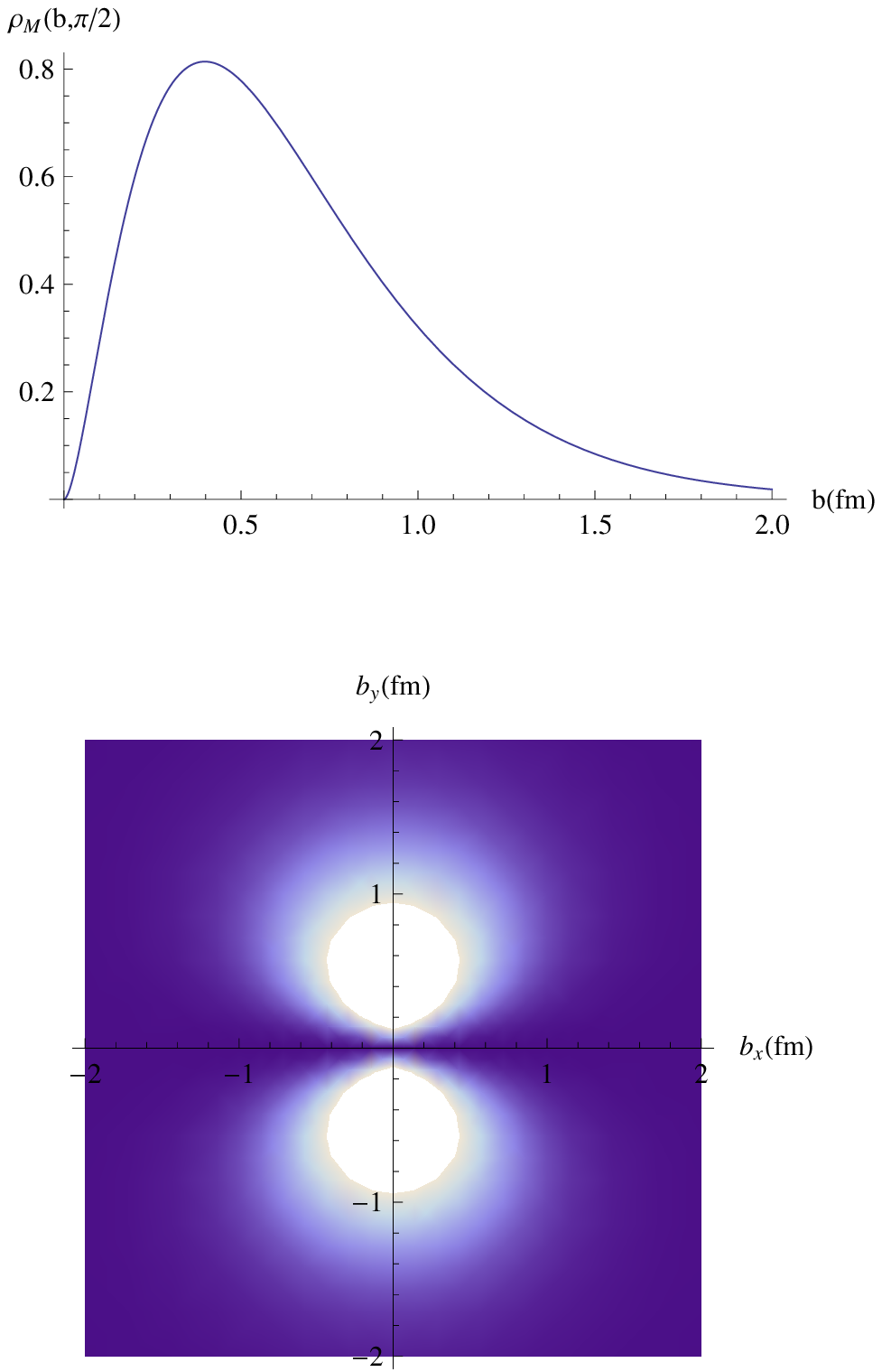}
\end{picture}
\caption{Upper panel: $\tilde{\rho}_M(\bfb=(b,\phi=\pi/2))$ as a function of $b$.  Lower panel: Density plot of   $\tilde{\rho}_M(\bfb)$ .
The horizontal axis is the direction of the applied magnetic
field. The largest (smallest) values of   $\tilde{\rho}_M$ are denoted by the
brightest (darkest) areas. This figure is obtained using a dipole
parametrization for $F_2$ of the proton. }
\end{figure}

\subsection{Non-cylindrically symmetric  transverse charge density and  
shape of the nucleon}
\begin{figure}[ht]
\begin{center}
\includegraphics[width =7.7cm]{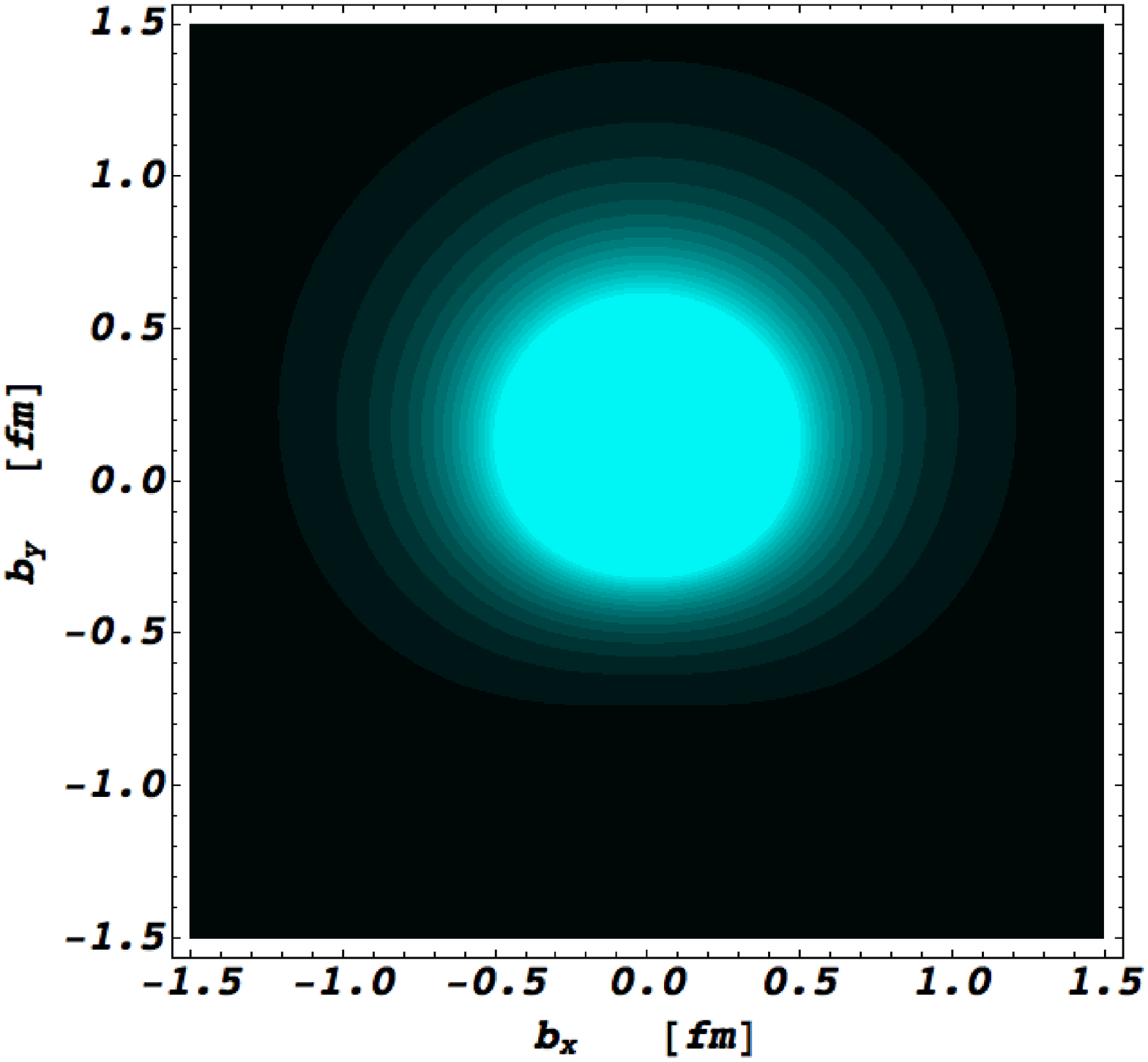}
\end{center}
\hspace{.5cm}
\includegraphics[width =7.5cm]{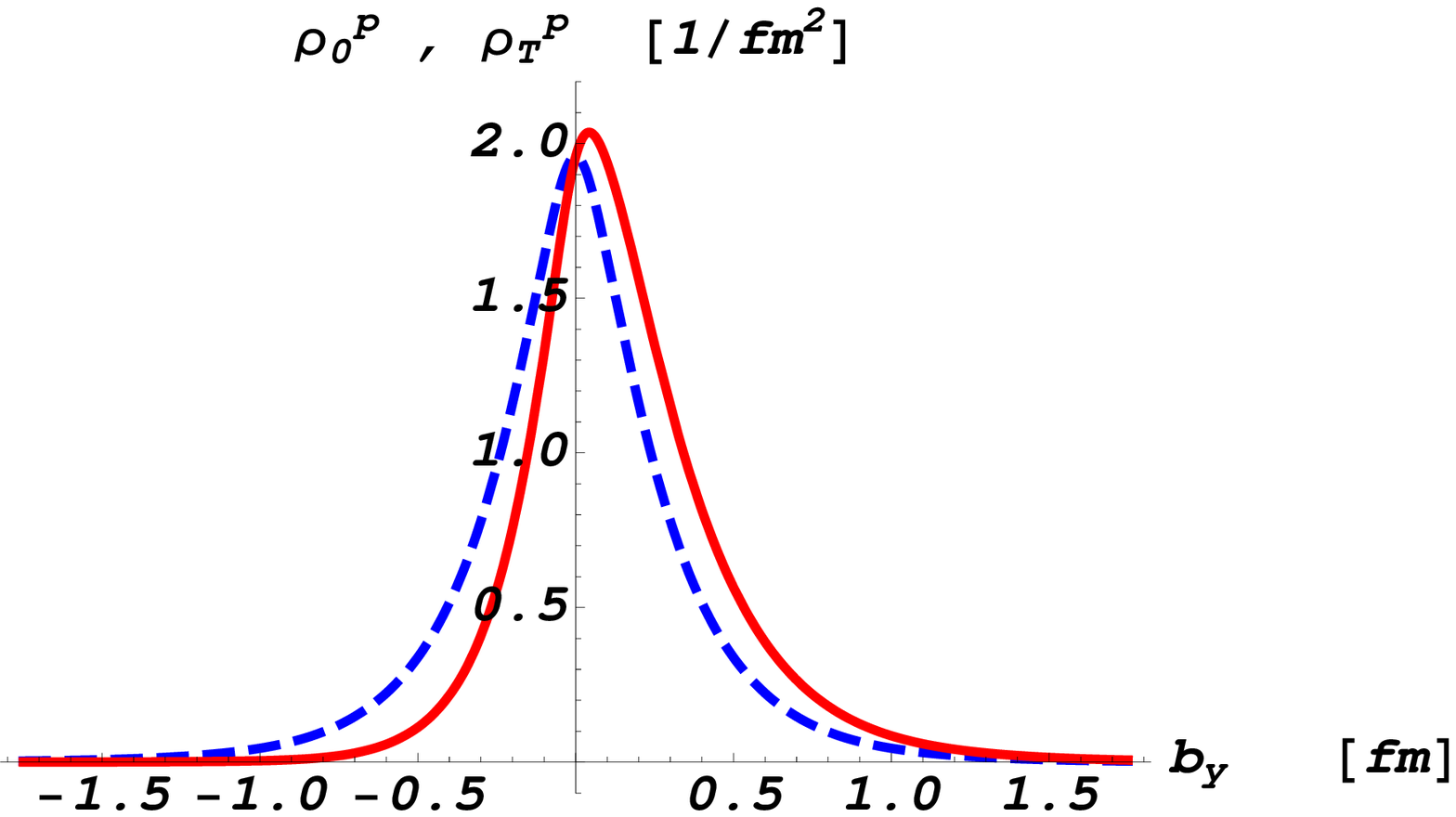}
\caption{Quark transverse charge densities in the  proton, after 
\cite{Carlson:2007xd}, figure provided courtesy of M.  Vanderhaeghen. 
The upper panel shows the density in the transverse plane for a 
proton polarized along the $x$-axis. The light (dark) regions correspond with
largest (smallest) values of the density. 
The lower panel compares the density along the $y$-axis 
for an unpolarized proton (dashed curve), 
and for a proton polarized along the $x$-axis (solid curve). 
The proton  form factors are from 
Arrington {\it et al.}~\cite{Arrington:2007ux}.  The momentum
transfer \bfq is the momentum added to the target proton.}
\label{figure8}
\end{figure}

Carlson and Vanderhaeghen \cite{Carlson:2007xd} computed the
transverse charge density$\rho^N_T(\bfb) $ in a given state of
transverse polarization 
For the case that  
the polarization is in the $x$ direction their result is expressed 
 as 
\bea
\rho^N_T(\bfb) =\rho(b)+\sin\phi\int_0^\infty
    {dQ\over 2\pi}{Q^2\over 2M}J_1(bQ)F_2(Q^2), \label{cv}
\eea
where $\rho(b) $ is obtained from \eq{rhoblong}. The sign of the second
term of \eq{cv} is obtained using the definition that $\bfq$ is the
momentum absorbed by the target nucleon. Their results for the proton
are shown in Fig.~\ref{figure8}.

The magnetization density $\tilde{\rho}_M(\bfb)$ presented here and
the transverse charge density of Ref.~\cite{Carlson:2007xd}  
each depend on the direction of $\bfb$, thus indicating a
non-spherical shape of the nucleon or the violation of cylindrical
symmetry. These violations of cylindrical symmetry should not be
confused with those present in earlier papers on the 
shape of the nucleon 
Refs.~\cite{Miller:2003sa,Kvinikhidze:2006ty,Miller:2007ae} that
define a shape via the matrix elements of a spin-dependent density
operator.
 The
transverse 
charge densities that are the main subject of this article are matrix
elements
 of currents that are local in transverse coordinate space and  are
 related to
 GPDs. In contrast, those shapes of 
\cite{Miller:2003sa,Kvinikhidze:2006ty,Miller:2007ae}, computed from
momentum-space wave functions and probabilities are matrix
elements of
 quark densities for given values of transverse momenta and are 
related to TMDs. It should also be noted that the first example of
\cite{Miller:2003sa} is presented in coordinate space. The infinite
momentum frame version of the spin-dependent density operator is 
$q_+^\dagger(0,\bfb) {1\over2}(1+{\bf n}\cdot\boldgamma)q_+(0,\bfb)$,
where ${\bf n}$ is an arbitrary transverse direction
\cite{Diehl:2005jf}. Evaluating the matrix element of this quantity
produces the spin-dependent density, which can be thought of as 
the $x^-$ integrated version of the coordinate space results of
 \cite{Miller:2003sa}.
The proton is non-spherical if the function $\tilde{A}''_{T10}$ of
\cite{Gockeler:2006zu} is non-vanishing, as recent lattice
calculations show \cite{Schierholz:2009xx}.
One can also examine several different
generalized densities $q_+^\dagger(0,\bfb) \Gamma   q_+(0,\bfb)$, 
with operators $\Gamma$ denoted in
\cite{Diehl:2005jf}.
\section{TRANSVERSE TRANSITION CHARGE DENSITIES}
There is much experimental interest in measuring transition 
form factors that represent 
 probability amplitudes for a nucleon undergoing  electron scattering
to make a transition to a given baryon resonance. Substantial
experimental efforts are underway at electron-scattering 
laboratories such as Jefferson
Laboratory, 
ELSA in Bonn, and MAMI in Mainz.  Moreover, there is considerable
interest in using lattice techniques to make QCD
calculations of these transition amplitudes.

Carlson \& Vanderhaeghen \cite{Carlson:2007xd} used empirical
information regarding the $N\to\Delta$ transition form factors to map
out the transition charge density.
This, in a transversely polarized N and 
$\Delta$, contains monopole, dipole and quadrupole patterns.
 The latter corresponds to  a deformation of the nucleon 
and $\Delta$ transverse charge density. Substantial deformations are observed.
Lattice QCD calculations have been applied to these very same
transition densities
\cite{Alexandrou:2008bn,Alexandrou:2009hs}. The $\Delta^+$ charge density
is found to be elongated along the axis of the spin and the quadrupole
moment is larger than the value characterizing a point particle. This
means the the $\Delta^+$ is prolately deformed.

Tiator \& Vanderhaeghen \cite{Tiator:2008kd}
used recent experimental data to 
 analyze the electromagnetic transition from the nucleon to the
$ P_{11}(1440) $
 resonance, which is often believed 
to be the first radial excitation of the proton.
 They used the empirical transition form factors to 
  find 
  that the transition from the proton to  the 
$ P_{11}(1440) $
is dominated by up quarks in a central region of width of about  0.5
fm 
and by down quarks in an outer band which extends up to about 1 fm. 
Tiator {\it et al.} \cite{Tiator:2009mt}
are extending the study of transition transverse
densities to  the $S_{11}$ and $D_{13}$ resonances.

\section{NUCLEAR TRANSVERSE CHARGE DENSITIES}

The very same infinite momentum frame formalism can be applied to the
analysis of nuclear form factors. The non-relativistic approximation
of \eq{nrcond} is generally applicable for existing measurements of 
nuclear form factors \cite{Miller:2009sg}. However, meson exchange
currents are known to be important for $Q^2>2 $ GeV$^2$
\cite{Riska:1979hz}. In 
 that  case, the charged constituents have
different masses (see \eq{nrdensi}), the pion moves
relativistically, and
  the charge density is not a
three-dimensional 
Fourier transform of the electromagnetic form factor. Moreover,
Jefferson Laboratory plans to considerably 
extend the available range of momentum transfer for light nuclear form
factors.
Hence transverse charge densities  which supply model independent
information are of considerable relevance for  nuclear physics.

Carlson \&  Vanderhaeghen \cite{Carlson:2008zc} made the first
computation of a nuclear
transverse density  in their analysis of  the deuteron empirical
transverse charge densities \cite{Abbott:2000ak}.  The 
charge densities are  characterized by monopole, dipole and
quadrupole patterns because the deuteron has a spin of unity.  
Ref.~\cite{Carlson:2008zc} observes
 a dip in the center-of-charge
density for helicity-zero deuterons. This   central depression is in
accord  with standard nuclear force model calculations \cite{Forest:1996kp}
and is 
 a simple consequence of the deuteron $D$-state.
 We note that the deuteron is a non-relativistic
system in the sense defined by Ref.~\cite{Miller:2009sg}. Thus these
densities can, with a high degree of accuracy, be thought of as an
integral over $z$ of the corresponding non-relativistic density; see
\eq{zint}. The dip is an important finding, because it
is model independent.
Ref.~\cite{Carlson:2008zc} also finds that 
transversely polarized deuterons show dipole
and quadrupole structure in the charge densities.
Their electric dipole and quadrupole moments only depend
on the spin-1 particle's anomalous magnetic dipole
moment and its anomalous electric quadrupole moment,
arising from its internal structure.

Nuclei are non-relativistic systems \cite{Miller:2009sg} so that for
presently available data the principle distinction between transverse
densities and the usual three-dimensional Fourier transform is that
the former is an integral \eq{zint} over the latter. However,
 existing nuclear
data extend only to about $Q^2=1$ GeV$^2$
\cite{Amroun:1994qj,Sick2001245,Sick:2008zza}. In this case the motion of
the nucleons within the nucleus is just barely relativistic. If, as
expected, data are taken at higher values of $Q^2$ relativistic effects
can be expected to be important. One example is the nuclei with
$A=3$ and mass $M_3$ and Sachs electric and magnetic form factors
$F_C,F_M$ \cite{Amroun:1994qj}.
If $Q^2/4M_3^2$ is of order unity, the 
distinction between $F_1 =F_C-Q^2/4M_3^2F_M$ and $F_C$ will become
important and the  role of transverse densities may become very important.

\subsection{Other Applications}

 Transverse momentum densities, giving the momentum density of
 hadrons, have also been studied \cite{Abidin:2008sb,Selyugin:2009ic}.
The spatial distribution of the momentum component $P^+$ is found to
 be related to Fourier transforms of gravitational form factors
 \cite{Abidin:2008sb}
which in turn are related to experimentally observed
 data. This is because gravitational form factors  
  are matrix elements of the energy-momentum tensor obtainable as
 second Mellin moments of GPDs \cite{Ji:1996ek,Ji:1996nm,Abidin:2008ku}.
The proton momentum density in the transverse plane
 plane was found to be more compact than its charge density
 \cite{Abidin:2008sb}.

Any charge density will deform when subjected to an external electric  field;
related measurable quantities are called polarizabilities. 
Gorchtein {\it et al.} \cite{Gorchtein:2009qq} 
 extended the transverse charge density 
 formalism to extract light-front quark charge densities 
related to polarizabilities
and  showed that the resulting induced
polarizations can be
 extracted from proton generalized polarizabilities
\cite{Guichon:1995pu,Drechsel:1997xv}. The available
 data for the 
generalized electric polarizabilities of the proton yield a transverse
density that shows an usual oscillatory pattern.

\section{SUMMARY AND FUTURE DIRECTIONS}
Transverse  charge densities are a new tool for  analyzing
electromagnetic form factors of systems composed of constituents that
move relativistically. These quantities are hadronic charge
densities as seen in a reference frame 
  moving  with infinite momentum. One of the
main advantages of the infinite momentum framework is that boosts in
the transverse direction form
a kinematic  sub-group of the \poinc group, so that 
transformations to frames moving in a direction transverse  to that of
the infinite momentum are  carried out using the  transverse position
operator. This is just like the usual
non-relativistic Galilean transformation. Thus the two-dimensional
Fourier transformation of electromagnetic form factors provides a
rigorous way to study charge and magnetization densities.
Transverse charge densities involve matrix elements of local operators 
$q_+^\dagger(t=0,z=0,\bfb) \Gamma q_+(t=0,z=0,\bfb)$, computable  using
any of the
light-front, equal-time or lattice techniques.

The use of transverse momentum densities has led to some
very interesting findings. Examples include the negative
nature of   the central neutron charge density
 \cite{Miller:2007uy} as caused by the dominance of $d$ quarks at the
 center \cite{Miller:2008jc}, the seemingly singular nature of the 
density at the center of the pion
\cite{Miller:2009qu}, the spatial extent of the proton's magnetization 
density is greater than that of 
its charge density  \cite{Miller:2007kt}, 
and the deformation of the nucleon
\cite{Miller:2003sa,Schierholz:2009xx} 
  and  
the $\Delta$ baryon
\cite{Carlson:2007xd}  is substantial. 
The positive sign of the electron anomalous magnetic moment is
 explained \cite{Hoyer:2009sg}.
 
Transverse densities 
 have been used to analyze other baryon transitions
\cite{Tiator:2008kd}, 
nuclear charge densities \cite{Carlson:2008zc}
momentum densities \cite{Abidin:2008ku} and generalized
polarizabilities \cite{Gorchtein:2009qq}. 

Measurements of form factors at larger values of $Q^2$ than are
presently available   is needed to test
the singularity of the pionic transverse density and the inner core of
the neutron transverse density.
It also seems likely  that the use of transverse charge
densities will become the chosen tool to analyze nuclear charge
distributions once very high-momentum transfer data become available.  
Transverse
densities are relevant whenever matrix elements of an operator
$q_+^\dagger(t=0,z=0,\bfb) \Gamma q_+(t=0,z=0,\bfb)$
appears. Thus one may expect  to
see a host of interesting, informative and
important future applications of transverse densities.

\section{ACKNOWLEDGMENTS}

This work is partially supported by the USDOE. I thank 
J. Arrington, A. Bernstein,
 M. Burkardt, C. Carlson,  I. Clo\"{e}t, P. Hoyer, 
 E. Piasetzky, G. Ron, A. Radyushkin, J. Rinehimer,
B. Tiburzi, M. Vanderhaeghen and L. Zhu for useful discussions
regarding 
 transverse charge distributions.
\bibliographystyle{arnuke_revised.bst}
\bibliography{transverse}
\end{document}